\documentclass[prd,tightenlines,11pt,a4paper,onecolumn,preprintnumbers,superscriptaddress,nofootinbib]{revtex4}
\pdfoutput=1
\usepackage{amsmath}
\usepackage{amssymb}
\usepackage{graphicx}
\usepackage{xspace}
\usepackage{color}
\usepackage{slashed}
\usepackage{units}
\usepackage[hyperfootnotes=false]{hyperref}
\usepackage[dvipsnames]{xcolor}

\def \L {\mathcal{L}} %Lagrangian density

\def \vec#1{{\boldsymbol{#1}}}
\newcommand{\matrixx}[1]{\begin{pmatrix} #1 \end{pmatrix}} %Matrix with brackets
\newcommand{\hc}{\ensuremath{\text{h.c.}}}
\newcommand{\dd}{\mathrm{d}}
\newcommand{\tr}{\mathrm{tr}}
\newcommand{\diag}{\mathrm{diag}}

\makeatletter
\newcommand\footnoteref[1]{\protected@xdef\@thefnmark{\ref{#1}}\@footnotemark}
\makeatother

\begin{document}

\title{Pati--Salam explanations of the \texorpdfstring{$B$}{B}-meson anomalies}

\preprint{ULB-TH/18-11, UCI-TR-2018-13}

\author{Julian Heeck}
\email{julian.heeck@uci.edu}
\affiliation{Service de Physique Th\'eorique, Universit\'e Libre de Bruxelles, Boulevard du Triomphe, CP225, 1050 Brussels, Belgium}

\affiliation{Department of Physics and Astronomy, University of California, Irvine, CA 92697-4575, USA}

\author{Daniele Teresi}
\email{daniele.teresi@df.unipi.it}
\affiliation{Service de Physique Th\'eorique, Universit\'e Libre de Bruxelles, Boulevard du Triomphe, CP225, 1050 Brussels, Belgium}

\affiliation{Dipartimento di Fisica ``E. Fermi'', Universit\`a di Pisa,
Largo Bruno Pontecorvo 3, I-56127 Pisa, Italy}

\hypersetup{
    pdftitle={
		Pati--Salam explanations of the \texorpdfstring{$B$}{B}-meson anomalies},
    pdfauthor={Julian Heeck, Daniele Teresi}
}

\begin{abstract}

We provide a combined explanation of the increasingly tantalizing $B$-meson anomalies, both in $R_{K^{(*)}}$ and $R_{D^{(*)}}$, in the Pati--Salam model with minimal matter content. This well-known model, based on the gauge group $SU(4)_{LC}\times SU(2)_L\times SU(2)_R$, naturally contains a variety of scalar leptoquarks with related and restricted couplings. In particular we show that the seesaw-motivated scalar leptoquark within the representation $(\vec{\overline{10}},\vec{3},\vec{1}) $ and its right-handed parity partner $(\vec{\overline{10}},\vec{1},\vec{3}) $ can solve both anomalies while making testable predictions for related observables such as $B\to K\nu\nu$ and $B\to K \mu\tau$. The solution of the $R_{K^{(*)}}$ anomaly alone can be related to a type-II seesaw neutrino mass structure. Explaining also $R_{D^{(*)}}$ requires the existence of a light right-handed neutrino, which constrains the UV structure of the model.

\end{abstract}

\maketitle

\section{Introduction}
In the last few years various hints for new physics manifesting itself in lepton flavour universality violation (LFUV) in semileptonic $B$-meson decays have been reported. Several experimental results by different collaborations point at an increasingly coherent picture. 

A first set of anomalies involves the charged-current transition $b \to c \tau \nu$, with LFUV between the $\tau$ and the other charged leptons in the decay $B \to D^{(*)} \ell \nu$, as parametrized by the double ratio $R_{D^{(*)}}$ with respect to the Standard Model (SM) prediction.  This observable has been measured by the BaBar~\cite{Lees:2012xj,Lees:2013uzd}, Belle~\cite{Huschle:2015rga,Sato:2016svk,Hirose:2016wfn}, and LHCb~\cite{Aaij:2015yra,Aaij:2017uff,Aaij:2017deq} collaborations and is found to be consistently above the SM predictions.
Global fits~\cite{HFLAV:2018} give a combined statistical evidence at the $\sim 4\sigma$ level. 
A second set of anomalies appears in the neutral-current transition $b \to s \mu \mu$, with hints for LFUV in the decays $B \to K^{(*)} \ell \ell$~\cite{Aaij:2014ora,Aaij:2017vbb}, parametrized by the double ratio $R_{K^{(*)}}$ between the $\mu$ and $e$ channels, relative to the SM. This is further corroborated by a deviation measured in the angular distribution of the process $B\to K^{*}\mu^+\mu^-$~\cite{Aaij:2015oid,Aaij:2013qta}, for which the theoretical SM prediction is however still debated. Again, global fits give a deviation from the SM at the $\sim 4 \sigma$ level or even higher~(see e.g.~Refs.~\cite{Capdevila:2017bsm,Altmannshofer:2017yso,Geng:2017svp,Ciuchini:2017mik,DAmico:2017mtc,Hiller:2017bzc}).

Finding combined explanations in terms of physics beyond the SM is far from trivial, mainly because of the large new-physics effects needed to explain $R_{D^{(*)}}$ that have to compete with the tree-level SM contribution. The effective-field-theory (EFT) scale involved is rather low at $\Lambda \simeq \unit[3.4]{TeV}$, and therefore a number of related observables pose stringent constraints; one of the most important observables here is $B \to K \nu \nu$ because the involved quark-level transition $b \to s \nu \nu$ is often related to the same new physics that contributes to the anomalous $b \to c \tau \nu$ by $SU(2)_L$ invariance. 
Explaining $R_{K^{(*)}}$ by itself is instead much easier since the EFT scale involved is $\Lambda \simeq \unit[31]{TeV}$. At the level of simplified models, combined explanations for $R_{K^{(*)}}$ and $R_{D^{(*)}}$ involving only left-handed new physics~\cite{Buttazzo:2017ixm} make use of a single vector leptoquark (LQ) $U_\mu$ with SM $(SU(3)_C,SU(2)_L)_{U(1)_Y}$ quantum numbers $ (\vec{3}, \vec{1})_{2/3}$, since this particle does not contribute to $B \to K \nu \nu$. With more than one new multiplet one can for example consider the combination of the scalar LQs $S_1=(\vec{3}, \vec{1})_{1/3}$ and $S_3=(\vec{3}, \vec{3})_{1/3}$, both with left-handed couplings, which can interfere destructively in the dangerous observables~\cite{Crivellin:2017zlb,Buttazzo:2017ixm}. Alternatively, simplified models explaining the charged-current anomaly by \emph{right-handed} new physics by introducing a light right-handed neutrino that plays the role of the active ones in $R_{D^{(*)}}$ typically do not run into problems with $B \to K \nu \nu$~\cite{Becirevic:2016yqi,Asadi:2018wea,Greljo:2018ogz,Robinson:2018gza,Azatov:2018kzb} because $b \to s \nu \nu$ is no longer the $SU(2)_L$ conjugate of the anomalous transition.

Recently, more and more effort has been put attempting to go beyond simplified models and  consider instead full models for the combined explanation of both sets of anomalies. In this respect, lepton-colour unification as given by the Pati--Salam (PS) symmetry $SU(4)_{LC} \times SU(2)_L \times SU(2)_R$ has received a lot of attention because this is arguably the most natural framework to embed LQs in a complete model.\footnote{In addition, we also mention very recent solutions based on lepton-quark unification different from PS~\cite{Becirevic:2018afm,Faber:2018qon}.} 
The obvious choice is to try to explain the anomalies with the vector LQ $U_\mu$~\cite{Assad:2017iib}, since it is contained in the set of PS gauge bosons. However, since the anomalies point at new physics in the TeV range, such a light PS gauge boson would conflict with experiments involving first-generation fermions, for instance $K_L \to e^\pm \mu^\mp$, which exclude PS scales lower than $\sim \unit[1000]{TeV}$~\cite{Valencia:1994cj,Smirnov:2007hv,Kuznetsov:2012ai,Smirnov:2018ske}. A number of modifications to the PS model have been suggested to circumvent this problem, in particular by suppressing the coupling of the gauge boson $U_\mu$ to the first generation in different ways: i) introducing many vector-like fermions that mix with the SM ones and generate flavour-dependent gauge couplings~\cite{Calibbi:2017qbu,DiLuzio:2017vat}, ii) placing the PS symmetry in a Randall--Sundrum  background to suppress the dangerous observables by a conveniently chosen localization in the extra dimension~\cite{Blanke:2018sro}, iii) introducing three copies of the PS symmetry, one for each generation~\cite{Bordone:2017bld}, iv) giving up the gauge character of the PS symmetry by considering it as a global symmetry of a strongly-interacting theory~\cite{Barbieri:2017tuq}, which can also relate the $B$-meson anomalies to a possible solution of the gauge hierarchy problem, v) more exotic options~\cite{Aydemir:2018cbb}.

While these are all perfectly viable possibilities, each with its own advantages, they often involve ad-hoc modifications of the Pati--Salam framework. This is mainly due to the usage of the vector LQ to explain the anomalies. However, the PS model also naturally contains \emph{scalar} LQs, which have flavour-\emph{dependent} couplings; the first-generation LQ couplings can therefore be relatively suppressed without any model-building effort. The PS model is typically (but not necessarily) supplemented by a left--right exchange symmetry, often in the form of parity. In this case, as we are going to show in this work, a \emph{single} representation of the parity-symmetric PS group,
\begin{equation}
(\overline{\vec{10}},\vec{3},\vec{1}) \oplus (\overline{\vec{10}},\vec{1},\vec{3}) \,,
\end{equation}
typically already present in the PS model with minimal matter content, is sufficient to explain both sets of anomalies, basically without any tension with other observables. This representation contains two scalar LQs that can be used to explain the anomalies, one with left-handed and one with right-handed couplings. However, it is far from trivial that a successful explanation is possible, because the PS symmetry restricts the structure of the couplings (quarks and leptons have the same couplings), parity makes the left-handed and right-handed ones equal, and their different chirality forbids them from interfering. The required suppression of $B \to K \nu \nu$ can therefore not be achieved by the destructive interference between several LQs, as discussed above, but instead requires one of the right-handed neutrinos to be light with a particular mixing pattern.

The plan of the paper is the following: in Sec.~\ref{sec:PS} we introduce the PS model with minimal matter content, and discuss the restrictions on the couplings posed by its symmetries. In Sec.~\ref{sec:RK} we show that explaining the neutral-current anomalies alone can be related to a type-II seesaw mass structure for the active neutrinos, leading to testable predictions. We then move to the combined explanation, by first discussing the model-building challenges involved in Sec.~\ref{sec:challenges} and then presenting the combined explanation, together with its phenomenological predictions, in Sec.~\ref{sec:combined}. We finally conclude in Sec.~\ref{sec:conclusion} and report the details of the fits for the explanations of the anomalies in App.~\ref{app:RK} and~\ref{app:RD}.
We discuss the renormalization group running of the LQ and gauge couplings in App.~\ref{app:RGE} and App.~\ref{app:RG_gauge}, respectively, to illustrate the validity of our assumed structures.

\section{Pati--Salam model}\label{sec:PS}

The Pati--Salam model is based on the gauge group $SU(4)_{LC}\times SU(2)_L\times SU(2)_R$~\cite{Pati:1974yy}, where colour and lepton number are unified in a single $SU(4)_{LC}$ factor. The PS group is eventually broken to the SM gauge group $SU(3)_C\times SU(2)_L \times U(1)_Y$, for which we define the hypercharge $Y=Q-T_L^3$.
The SM fermions (plus right-handed neutrinos) are minimally unified in the following PS representations,
\begin{align}
\Psi_L \sim (\vec{4},\vec{2},\vec{1}) &\to \left(\vec{3},\vec{2}\right)_{\tfrac{1}{6}}\oplus\left(\vec{1},\vec{2}\right)_{-\tfrac{1}{2}} \equiv Q_L \oplus L_L\,,\\
\Psi_R \sim (\vec{4},\vec{1},\vec{2}) &\to \left(\vec{3},\vec{1}\right)_{\tfrac{2}{3}}\oplus \left(\vec{3},\vec{1}\right)_{-\tfrac{1}{3}}\oplus \left(\vec{1},\vec{1}\right)_{-1}\oplus \left(\vec{1},\vec{1}\right)_{0} \equiv u_R \oplus d_R\oplus \ell_R\oplus N_R\,,
\end{align}
where we also give the branching rules with regards to the SM and identify the components with their usual SM notation, i.e.~left-handed quark doublet $Q_L$ etc, suppressing generation indices.
In the presence of only this minimal matter content, fermion masses can be generated by introducing complex scalars in the representations in $(\vec{\overline{4}},\vec{2},\vec{1})\otimes (\vec{4},\vec{1},\vec{2})=(\vec{1},\vec{2},\vec{2}) \oplus (\vec{15},\vec{2},\vec{2})$~\cite{Pati:1983zp}:
\begin{align}
\phi_\vec{1} \sim (\vec{1},\vec{2},\vec{2}) &\to \left(\vec{1},\vec{2}\right)_{\tfrac{1}{2}}\oplus\left(\vec{1},\vec{2}\right)_{-\tfrac{1}{2}} ,\\
\phi_\vec{15} \sim (\vec{15},\vec{2},\vec{2}) &\to \left(\vec{1},\vec{2}\right)_{\tfrac{1}{2}}\oplus \left(\vec{1},\vec{2}\right)_{-\tfrac{1}{2}}\oplus \left(\vec{3},\vec{2}\right)_{\tfrac{1}{6}}\oplus \left(\vec{\overline{3}},\vec{2}\right)_{-\tfrac{1}{6}}\notag\\
&\quad\oplus \left(\vec{3},\vec{2}\right)_{\tfrac{7}{6}}\oplus \left(\vec{\overline{3}},\vec{2}\right)_{-\tfrac{7}{6}}\oplus \left(\vec{8},\vec{2}\right)_{\tfrac{1}{2}}\oplus \left(\vec{8},\vec{2}\right)_{-\tfrac{1}{2}},
\end{align}
which have Yukawa couplings
\begin{align}
\L \supset \overline{\Psi}_L \left( k_{\vec{1},A} \phi_\vec{1}+ k_{\vec{1},B} \phi_\vec{1}^c +k_{\vec{15},A} \phi_\vec{15}+ k_{\vec{15},B} \phi_\vec{15}^c \right) \Psi_R +\hc ,
\end{align}
where the superscript $c$ denotes an appropriate form of charge conjugation and we suppress all indices.
$\phi_\vec{1}$ and $\phi_\vec{15}$ contain two electroweak doublets each and can thus be used to both break $SU(2)_L\times U(1)_Y\to U(1)_Q$ and generate fermion masses. $\phi_\vec{1}$ by itself would preserve the $SU(4)_{LC}$ multiplet and thus lead to the mass-unification relations~\cite{Pati:1983zp,Volkas:1995yn} 
\begin{align}
m_d = m_\ell \,, \quad m_u = m_N^\text{Dirac}
\label{eq:mass_unification}
\end{align}
for the Dirac mass matrices of down quarks with charged leptons, and up quarks with Dirac neutrinos. The latter relation will be broken by the seesaw mechanism, as discussed below, leaving $m_d = m_\ell$ as a PS prediction. Taking into account that these relations hold at the potentially high PS scale and that renormalization-group running strongly affects quarks, this is a potentially viable relation~\cite{Volkas:1995yn}.
Keeping also the $\phi_\vec{15}$ vacuum expectation values (VEVs) allows one to break the relation of Eq.~\eqref{eq:mass_unification} at tree level and generate arbitrary masses for all fermions~\cite{Pati:1983zp}. Since the bottom--tau unification seems to be a good approximation at least, we will assume that the $\phi_\vec{15}$ VEVs only lead to small perturbations. As a result, the diagonalization of the mass matrices will be approximately the same for quarks and leptons, so diagonalization will proceed through
\begin{align}
u_{L,R} \to V^\dagger_{L,R} u_{L,R} \,, \quad \nu_L \to V^\dagger_L \nu_L \,,\quad N_R \to V^\dagger_R N_R \,,
\end{align}
where we chose to pick the down-quark and charged-lepton basis without loss of generality. $V_L$ is the usual unitary Cabibbo--Kobayashi--Maskawa (CKM) mixing matrix and $V_R$ the right-handed analogue, which is physical due to the additional interactions beyond the SM. Imposing furthermore parity $\mathcal{P}$ on our theory, i.e.~$f_L \leftrightarrow f_R$, leads to a relation between the two mixing matrices, approximately setting them equal if we ignore phases: $V_R\simeq V_L$~\cite{Maiezza:2010ic,Senjanovic:2014pva}.
We will mostly employ the above relations in the following, but it should be kept in mind that these are very strong assumptions that could easily be loosened, leading to more freedom in the couplings.
Using generalized charge conjugation $\mathcal{C}$ instead of parity will lead to qualitatively similar results and will not be discussed further.

We still need additional scalars to break PS to the SM and ideally also for neutrino mass generation. A common choice to accomplish both tasks is given by the scalar representations~\cite{Pati:1983zp}
\begin{align}
\Delta_L \sim (\vec{\overline{10}},\vec{3},\vec{1}) 
&\to 
\left(\vec{6},\vec{3}\right)_{-\tfrac{1}{3}}\oplus
\left(\vec{\overline{3}},\vec{3}\right)_{\tfrac{1}{3}}\oplus
\left(\vec{1},\vec{3}\right)_{1} \equiv 
\Sigma_3 \oplus S_3 \oplus \delta_3\,,\\
\Delta_R \sim (\vec{\overline{10}},\vec{1},\vec{3}) 
&\to 
\left(\vec{6},\vec{1}\right)_{-\tfrac{1}{3}}\oplus 
\left(\vec{6},\vec{1}\right)_{\tfrac{2}{3}}\oplus 
\left(\vec{6},\vec{1}\right)_{-\tfrac{4}{3}} \oplus 
\left(\vec{\overline{3}},\vec{1}\right)_{\tfrac{1}{3}}\oplus 
\left(\vec{\overline{3}},\vec{1}\right)_{-\tfrac{2}{3}}\oplus 
\left(\vec{\overline{3}},\vec{1}\right)_{\tfrac{4}{3}} \notag\\ 
&\oplus 
\left(\vec{1},\vec{1}\right)_{0}\oplus  
\left(\vec{1},\vec{1}\right)_{1}\oplus  
\left(\vec{1},\vec{1}\right)_{2} \equiv 
\Sigma_1\oplus \tilde{\Sigma}_1\oplus\overline{\Sigma}_1\oplus
S_1\oplus \tilde{S}_1\oplus\overline{S}_1\oplus
\delta_1\oplus \tilde{\delta}_1\oplus\overline{\delta}_1\,,
\end{align}
where $\langle \delta_1 \rangle $ has the right quantum numbers for PS breaking and to generate Majorana masses for $N_R$. This VEV thus breaks the PS group down to the SM in one step and provides a mass $\propto \langle \delta_1 \rangle $ to the non-SM gauge bosons $X_\mu$ (the vector LQ), $W_{\mu,R}^+$, and $Z_\mu'$. As discussed below, the \emph{left-handed} triplet VEV $\langle \delta_3 \rangle$ can generate type-II seesaw~\cite{Magg:1980ut,Schechter:1980gr,Cheng:1980qt} masses for $\nu_L$ and is restricted to be below GeV from electroweak precision data~\cite{pdg}.  
Note that the bars and the tildes are part of the LQ names and do not  denote charge conjugation. Also notice that, as will be detailed below, these representations do not lead to dangerous proton decay~\cite{Mohapatra:1980qe}.

We have the Yukawa Lagrangian in the flavour basis, suppressing all indices,
\begin{align}
\L = \overline{\Psi}_L^c y^L \Delta_L \Psi_L + \overline{\Psi}_R^c y^R \Delta_R \Psi_R +\hc \,.
\end{align}
$y^{L,R}$ are \emph{symmetric} coupling matrices in flavour space, which is a true PS prediction. For instance, PS symmetry restricts the coupling of $S_3$ to a $b$ quark and a $\mu$ to be the same as to an $s$ quark and a~$\tau$; this has a large impact on the explanations of the anomalies and the related phenomenological predictions, as will be clear in the next sections.
Expanded, the couplings read
\begin{align}\label{eq:couplings_S3}
\overline{\Psi}_{L}^c y^L \Delta_L \Psi_{L} &\equiv 
6^{-1/4} \left(\overline{Q}_{L}^c y^L Q_{L} \Sigma_3\right)_{(\vec{1},\vec{1})_0}
+ \left(\overline{Q}_{L}^c y^L L_{L} S_3\right)_{(\vec{1},\vec{1})_0}
+ \frac{1}{\sqrt{2}} \left( \overline{L}_{L}^c y^L L_{L} \delta_3\right)_{(\vec{1},\vec{1})_0} \notag\\
&\supset \ \overline{\nu}_{L}^c  V_L^*  y^L  V_L^\dagger  u_{L} S_3^{-2/3}
-\frac{1}{\sqrt{2}} \left( \overline{d}_{L}^c y^L  V_L^\dagger  \nu_{L} + \overline{\ell}_{L}^c y^L V_L^\dagger u_{L}\right) S_3^{1/3}
+\overline{\ell}_{L}^c y^L d_{L} S_3^{4/3} \,,
\end{align}
using some Fierz identities and the fact that $y^L$ is symmetric. 
In the last line we give the couplings of the $S_3$ components $S_3^Q$ in the quark mass basis, where the superscript $Q$ denotes the electric charge.
For the right-handed fermions in their mass basis one finds
\begin{align}\label{eq:couplings_S1}
\overline{\Psi}_{R}^c y^R \Delta_R \Psi_{R} \equiv & 
- 2^{1/4}\,3^{-1/4}\, \overline{d}_{R}^c y^R V_R^\dagger u_{R}  \Sigma_1
+6^{-1/4} \, \overline{d}_{R}^c y^R d_{R}  \tilde{\Sigma}_1 
+6^{-1/4}\, \overline{u}_{R}^c V_R^* y^R V_R^\dagger u_{R} \overline{\Sigma}_1 \notag\\
&+  \overline{u}_{R}^c V_R^* y^R V_R^\dagger N_{R} \overline{S}_1 
-  \frac{1}{\sqrt{2}} \left(\overline{d}_{R}^c y^R V_R^\dagger N_{R} + \overline{\ell}_{R}^c y^R V_R^\dagger u_{R} \right)S_1
+  \overline{d}_{R}^c y^R \ell_{R} \tilde{S}_1 \notag\\
&+ \frac{1}{\sqrt{2}} \, \overline{\ell}_{R}^c y^R \ell_{R} \overline{\delta}_1
-  \overline{\ell}_{R}^c y^R V_R^\dagger N_{R} \tilde{\delta}_1
+ \frac{1}{\sqrt{2}}\, \overline{N}_{R}^c V_R^* y^R V_R^\dagger N_{R} \delta_1 \,.
\end{align}
We will be particularly interested in the couplings of the LQ $S_1$, given in the second line. Parity~$\mathcal{P}$ would exchange $\Psi_L \leftrightarrow \Psi_R$ and $\Delta_L \leftrightarrow \Delta_R$ and thus impose $y^L = y^R$ and $V_L=V_R$. These relations hold above the PS scale but are broken at lower energies, as can be seen by studying the renormalization group running of the couplings. As we show in App.~\ref{app:RGE} for some benchmark scenarios, the PS relations $y^R =y^L = (y^L)^\intercal$  at the TeV scale can still be satisfied to the $\mathcal{O}(10 \%)$ level, especially if the PS scale is not too high. For simplicity we will therefore work with the relation $y = y^\intercal \equiv y^L = y^R$ in the following. Again, it should be kept in mind that this is is a very constraining assumption that does not need to hold true, but we enforce it to maximize the predictivity of the framework. 
Note finally that within our minimal PS model the parity and PS breaking scales are identified so we must have the gauge-coupling relation $g_R = g_L$ at the PS scale. Given the low-energy particle content this fixes the PS scale to be typically around $\sim\unit[10^{13}]{GeV}$, as discussed in App.~\ref{app:RG_gauge}, although this can be changed by several orders of magnitude by the presence of states in the desert up to the PS scale.

One of the most interesting features of PS is the very existence of LQs, which we have already used in the naming scheme above.
We can identify the following scalar LQs within $\Delta_X$ in their common notation~\cite{Buchmuller:1986zs,Dorsner:2016wpm}: $S_3 = \left(\vec{\overline{3}},\vec{3}\right)_{\frac{1}{3}}$, $S_1 =\left(\vec{\overline{3}},\vec{1}\right)_{\frac{1}{3}}$, $\tilde{S}_1 =\left(\vec{\overline{3}},\vec{1}\right)_{\frac{4}{3}}$, and $\overline{S}_1 =\left(\vec{\overline{3}},\vec{1}\right)_{-\frac{2}{3}}$.\footnote{$\phi_\vec{15}$ also contains (non-chiral) scalar LQs, namely $R_2$ and $\tilde R_2$. The combinations $\tilde R_2+S_3$~\cite{Dorsner:2017ufx} and $R_2+S_3$~\cite{Becirevic:2018afm} have been discussed in relation to the $B$~anomalies and it has been shown that $R_2+S_3$ provides a good explanation. We will not discuss the LQs $R_2$ and $\tilde R_2$ in the following and simply assume them to be heavy.}
The PS origin of the $S_j$ LQs gives rise to many restrictions compared to naive SM extensions by $S_j$. For one, parity predicts right-handed partners to left-handed LQs. In particular, $S_3$ has a parity partner in the right-handed triplet $(S_1,\tilde{S}_1,\overline{S}_1)$. Below, we will use $S_3$ and $S_1$ to explain the $B$-meson anomalies. While this may look at first sight analogous to other existing approaches in the literature where scalar LQs with the same quantum numbers are used~\cite{Crivellin:2017zlb,Buttazzo:2017ixm}, our explanation to the $B$-meson anomalies is very different; our $S_1$ is forced to have purely right-handed couplings and  therefore cannot be used for a destructive interference with $S_3$ in $B \to K \nu \nu$ in order to suppress this channel. We will see that it is still possible to explain both $R_{K^{(*)}}$ and $R_{D^{(*)}}$ by invoking a \emph{light right-handed} neutrino, similar to Refs.~\cite{Becirevic:2016yqi,Asadi:2018wea,Greljo:2018ogz,Robinson:2018gza,Azatov:2018kzb}.

 Notice that the $S_j$ here are pure LQs without any di-quark couplings, as enforced by the PS symmetry. This has the immediate consequence that the above Yukawa couplings will not lead to proton decay, a welcome feature that is usually enforced by imposing rather ad-hoc baryon~\cite{Nieves:1981tv,Dorsner:2016wpm} or lepton-flavour symmetries~\cite{Hambye:2017qix}.\footnote{Our minimal PS model only has proton decay in the presence of $\phi_\vec{15}$ through some of the quartic interactions in the scalar potential~\cite{Pati:1983zp,Pati:1983jk}. This induces heavily suppressed dimension-10 proton-decay operators~\cite{ODonnell:1993kdg,Hambye:2017qix} that also depend on the quartic coupling and the Yukawas of $\phi_\vec{15}$ to the first generation, easy viable even for a low PS scale.}
Furthermore, the $S_j$ LQs here are purely chiral, every LQ only couples to one fermion chirality. This is again a welcome feature as it automatically leads to a chirality suppression in many rare decays and softens many constraints.

In Sec.~\ref{sec:RK} we will assume that the PS breaking to the SM is given entirely by $\langle \delta_1\rangle$, which also induces right-handed neutrino masses. If this is the only PS-breaking VEV, it needs to be above $\sim$ 1000 TeV to satisfy constraints on $K_L\to e^\pm \mu^\mp$ induced by the heavy coloured PS gauge boson~\cite{Valencia:1994cj,Smirnov:2007hv,Kuznetsov:2012ai,Smirnov:2018ske}. 
However, the key point of this work is that some of the PS \emph{scalars} can be lighter than this scale.\footnote{These scalars then suffer from the gauge hierarchy problem much like the Brout--Englert--Higgs boson in the SM supplemented by physical high-scale  thresholds. We will not attempt to solve this problem here.} 
Quartic couplings such as   $\text{Tr}\{\overline{\Delta}_X \Delta_X \overline{\Delta}_R \Delta_R\}$ will induce a mass splitting within the $\Delta_X$ multiplet proportional to $\langle \delta_1\rangle$. This makes it possible to have e.g.~$S_1$ at $\sim$ TeV but $\tilde{S}_1$ and $\overline{S}_1$ at a higher scale, so that we do not have to take their phenomenology into account in the analysis below.
We will assume for the most part that $S_3$ is around \unit[10]{TeV} and $S_1$ around \unit[1]{TeV} in order to address $R_{K^{(*)}}$ and $R_{D^{(*)}}$, respectively, while all other scalars are heavy enough to be ignored.\footnote{
The presence of LQs lighter than the PS scale will affect the gauge-coupling running and thus potential unification. Guaranteeing successful unification, e.g.~into $SO(10)$, requires the specification of the full particle spectrum. Since only particles up to few TeV are of phenomenological interest for the explanation of the anomalies, we will not specify the full spectrum and ignore gauge-coupling unification. However, partial unification of left and right gauge couplings, as required in our framework, is generically achieved as shown in App.~\ref{app:RG_gauge}.}

\section{Neutrino masses and \texorpdfstring{$R_{K^{(*)}}$}{R(K)}}\label{sec:RK}

Assuming that $\langle \delta_1 \rangle$ is the only PS-breaking VEV, our PS setup brings both type I and II seesaw contributions to the light Majorana neutrino masses, similar to standard left--right models~\cite{Akhmedov:2005np}:
\begin{align}\label{eq:seesaws}
M_\nu &= M_\nu^I + M_\nu^{II}\notag\\
 &\simeq - \frac{1}{\sqrt{2} \langle \delta_1\rangle} m_u \left(  V_R y^{R*} V_R^\intercal \right)^{-1} m_u - \sqrt{2} \langle \delta_3\rangle V_L^* y^L V_L^\dagger \,,
\end{align}
using Eq.~\eqref{eq:mass_unification} and the fact that the Dirac matrices are real and diagonal. Diagonalizing $M_\nu =U^* \text{diag} (m_{\nu_1},m_{\nu_2},m_{\nu_3}) U^\dagger$ with the usual unitary Pontecorvo--Maki--Nakagawa--Sakata (PMNS) mixing matrix $U$~\cite{pdg}, 
\begin{align}
U \equiv \matrixx{c_{12} c_{13} & s_{12} c_{13} & s_{13} e^{-i\delta}\\
	-c_{23} s_{12}- s_{23} s_{13} c_{12} e^{i\delta} & c_{23} c_{12}- s_{23} s_{13} s_{12} e^{i\delta} & s_{23} c_{13}\\
	s_{23}s_{12}- c_{23} s_{13} c_{12} e^{i\delta} & -s_{23} c_{12}- c_{23} s_{13} s_{12} e^{i\delta} & c_{23} c_{13}} \text{diag} (1, e^{i \alpha}, e^{i (\beta + \delta)}) \,,
\end{align}
allows us to predict the flavour structure of the coupling matrices $y^{L}=y^{R}$ using the measured neutrino masses and mixing angles (taken from Ref.~\cite{Esteban:2016qun}).

The neutral-current $B$-meson anomalies can all be explained by a new-physics Wilson-coefficient contribution $\Delta C_{9} = - \Delta C_{10} \simeq -0.61 \pm 0.12$~\cite{Capdevila:2017bsm} to the effective-Hamiltonian operator 
\begin{align}
\mathcal{O}_{9\mu} - \mathcal{O}_{10\mu} = - \frac{\alpha}{\pi v^2}\, V_{L,tb} V_{L,ts}^*  \, (\bar{s} \gamma_\mu P_L b)(\bar \mu \gamma^\mu  P_L \mu) \,,
\end{align}
where $\alpha$ is the fine-structure constant and $v\simeq \unit[246]{GeV}$ the electroweak VEV, both used purely for normalization.
It is well-known that this can be generated by the LQ $S_3 \in (\overline{\vec{10}},\vec{3},\vec{1})$ as
\begin{align}
\Delta C_9 = - \Delta C_{10} = \frac{\pi v^2}{\alpha} \,\frac{1}{V_{L,tb} V_{L,ts}^*} \, \frac{1}{M_{S_3}^2}  \, y^L_{23} y^{L*}_{22} \,,
\label{eq:C9}
\end{align}
with $\mathrm{Re}(\Delta C_9) \stackrel{!}{\simeq} -0.61 \pm 0.12$. In principle, a successful contribution can be easily realized with $M_{S_3} \simeq \unit[30]{TeV}$ for $y_{23}^L,y_{22}^L = \mathcal{O}(1)$. However, the PS symmetry and the neutrino oscillation data pose stringent constraints on the structure of the couplings. Therefore, in this section we will aim at explaining $R_{K^{(*)}}$ in relation to the neutrino mass structure.

A combined explanation of also the charged-current anomaly $R_{D^{(*)}}$ is more involved and cannot be related directly to the neutrino mass structure. Hence, this will be postponed to the next sections (Secs.~\ref{sec:challenges} and~\ref{sec:combined}).

\subsection{Type II dominance}

\begin{table}[t]
\raisebox{0.4em}{\begin{tabular}{c|c}
Parameter & Best fit \\
\hline
$M_{S_3}$ & $\unit[11.4]{TeV}$\\
$m_{\nu_1}$ & $\unit[2.1]{meV}$\\
$\delta$ & $-0.2 \pi$\\
$\alpha$ & $0.55\pi$\\
$\beta$ & $0.95\pi$\\
\hline
\hline
$\chi^2$ & $2.45$\\
\end{tabular}} 
\hspace{10ex}
\begin{tabular}{c|c|c}
Observable & Best fit & Pull/bound \\
\hline
$\Delta C_9 = - \Delta C_{10}$ & $-0.62$ & $-0.1\sigma$\\
${}^K\!R^{e/\mu}$ & $2.8 \times 10^{-5}$ & $-1.1\sigma$\\
${}^\pi\!R^{e/\mu}$ & $-2.1 \times 10^{-7}$ & $+1.1\sigma$\\
$K_{\pi \nu \nu}$ & $1.67$ & $-0.4\sigma$\\
\hline
$\mathrm{BR}(\mu \to e)$ & $6.9 \times 10^{-13}$ & $7 \times 10^{-13}$\\
$\mathrm{BR}(\mu \to e \gamma)$ & $1.1 \times 10^{-13}$ & $4.2 \times 10^{-13}$\\
$\Gamma(K_L \to e^\pm \mu^\mp)$ & $\unit[3.2 \times 10^{-29}]{GeV}$ & $\unit[6.0\times 10^{-29}]{GeV}$
\end{tabular} 
\caption{Results of the fit for the neutral-current anomalies and type-II neutrino masses, fixing the type-II seesaw VEV $\langle \delta_3 \rangle = \unit[50]{meV}$. \label{tab:RK}}
\end{table}

The flavour structure of the couplings can be determined by Eq.~\eqref{eq:seesaws}, under the assumptions discussed above, especially if one of the two terms dominate. Let us consider first the case in which the type-II seesaw contribution $M_\nu^{II}$ dominates, for instance because $\langle \delta_1\rangle$ is around the Grand Unified Theory (GUT) scale. Then, Eq.~\eqref{eq:seesaws} gives
\begin{align}
y^L = -\frac{1}{\sqrt{2} \langle \delta_3 \rangle}	 V_L^\intercal U^* \text{diag} (m_{\nu_1},m_{\nu_2},m_{\nu_3}) U^\dagger V_L \,,
\label{eq:typeIIansatz}
\end{align}
where $U$ ($V_L$) is the standard PMNS (CKM) mixing matrix. Since we know the right-hand side with the exception of the phases within $U$ and the overall neutrino mass scale $m_{\nu_1}$, we can predict the flavour structure of $y^L$ rather well.\footnote{Note that Eq.~\eqref{eq:typeIIansatz} does not require left--right symmetry and can thus be obtained even in generalized PS models.}
Assuming normal mass ordering with vanishing lightest neutrino mass $m_{\nu_1}\simeq 0$ as well as vanishing PMNS phases, the flavour structure is simply
\begin{align}\label{eq:yL_structure}
y^L \propto \matrixx{0.05 & 0.06 & -0.10 \\ 0.06 & 1 & 0.74 \\ -0.10 & 0.74 & 0.97} ,
\end{align}
setting all other neutrino~\cite{Esteban:2016qun} and CKM parameters to their best-fit values.
For $S_3$ in the $\mathcal{O}(\unit[30]{TeV})$ range we can easily fit $R_{K^{(*)}}$; however, the small but non-zero couplings to first generation fermions (both quarks and leptons due to the PS symmetry) give important constraints that need to be satisfied, in particular lepton flavour violation (LFV). The most stringent observable turns out to be $\mu\to e $ conversion in nuclei, because of the first and second-generation couplings to both leptons and the quarks in the nucleons. The relevant branching ratio in $\mathrm{Au}_{79}^{197}$ is given approximately by~\cite{Dorsner:2016wpm} 
\begin{equation}\label{eq:mutoe_appr}
\mathrm{BR}(\mu \to e) \simeq 1 \times 10^{-6}\, \left(\frac{\unit[30]{TeV}}{M_{S_3}}\right)^4\, |2 \, y^L_{11} y^{L*}_{12} + V_{L,us} y^L_{11} y^{L*}_{22}|^2 \simeq 6 \times  10^{-8} \left(\frac{\unit[30]{TeV}}{M_{S_3}}\right)^4 \, |y^L_{11} y^{L*}_{22}|^2 ,
\end{equation}
which has to be smaller than~$7 \times 10^{-13}$ at $90\%$~{\rm C.L.}~\cite{Bertl:2006up}.
The full expressions for this and the other relevant observables are given in App.~\ref{app:RK}. We see that the coupling $y^L_{11}$ needs to be suppressed by a factor $\sim 14$ with respect to the estimate of Eq.~\eqref{eq:yL_structure}. This can be achieved by a moderate tuning of the phases in the PMNS matrix together with $m_{\nu_1}$, in complete analogy to texture zeros in the neutrino mass matrix~\cite{Araki:2012ip}. In this way, one can relate the explanation of $R_{K^{(*)}}$ with information on the unknown parameters in the neutrino sector, most importantly the lightest neutrino mass $m_{\nu_1}$ and the Dirac CP phase $\delta$.

\begin{figure}[t]
\includegraphics[scale=1]{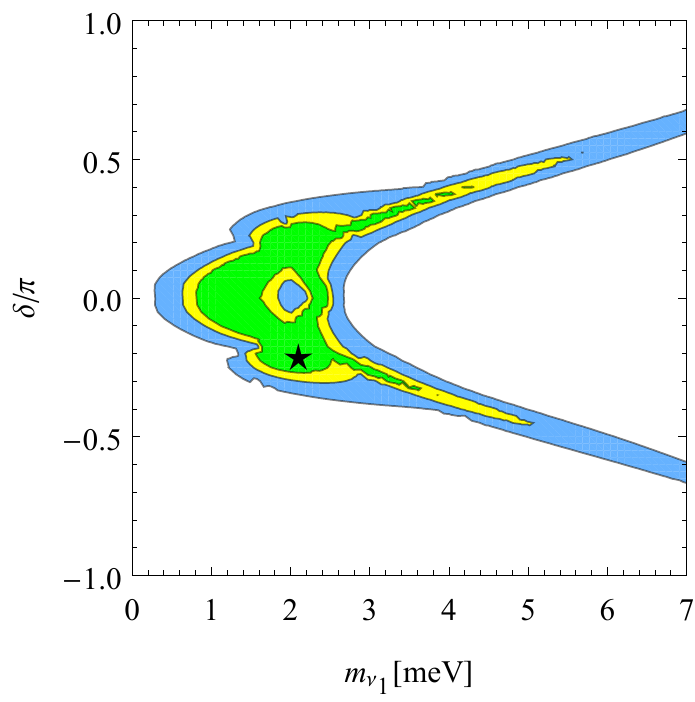}
\caption{
Results of the fit for the neutral-current anomalies and type-II neutrino masses, fixing $\langle \delta_3 \rangle = \unit[50]{meV}$. We show the $\Delta \chi^2 < 2.3$ ($1 \sigma$), $6.2$ ($2 \sigma$), $11.8$ ($3 \sigma$) regions in green, yellow, and blue, respectively, marginalising over all other parameters. The star denotes the best-fit point.
}
\label{fig:type_II}
\end{figure}

We perform a global fit including the most constraining observables discussed above together with other relevant ones involving first-generation fermions, detailed in App.~\ref{app:RK}. The results of the fit are given in Table~\ref{tab:RK}. The value of $M_{S_3}$ needed to explain $R_{K^{(*)}}$ depends of course on the absolute scale of the Yukawa matrix, which in turn depends on the unknown VEV $\langle \delta_3 \rangle$. Therefore, we cannot be particularly predictive on the scale of $M_{S_3}$ apart from the usual model-independent considerations based on perturbativity/unitarity~\cite{DiLuzio:2017chi}. Instead, a successful fit constrains the unknown neutrino-mass parameters, as discussed above. We show the two-dimensional correlation between $m_{\nu_1}$ and $\delta$ in Fig.~\ref{fig:type_II}. We note that the $1\sigma$ region for $\delta$ marginally overlaps with the $1\sigma$ range $\delta/\pi \simeq -0.7_{-0.17}^{+0.24}$ suggested by oscillation experiments~\cite{Capdevila:2017bsm}.

Considering the small number of free parameters it is remarkable that the Ansatz of Eq.~\eqref{eq:typeIIansatz} can indeed be used to resolve $R_{K^{(*)}}$. Ongoing and upcoming experimental efforts to significantly improve the limits on muonic LFV in $\mu$ to $e$ conversion (Mu2e~\cite{Bartoszek:2014mya}, COMET~\cite{Krikler:2015msn}), $\mu \to 3 e$ (Mu3e~\cite{Blondel:2013ia}), and $\mu\to e\gamma$ (MEG-II~\cite{Baldini:2018nnn}) will conclusively test this scenario, seeing as it is impossible to suppress all these observables much further.

\subsection{Type I dominance}

\begin{figure}[t]
\includegraphics[scale=1]{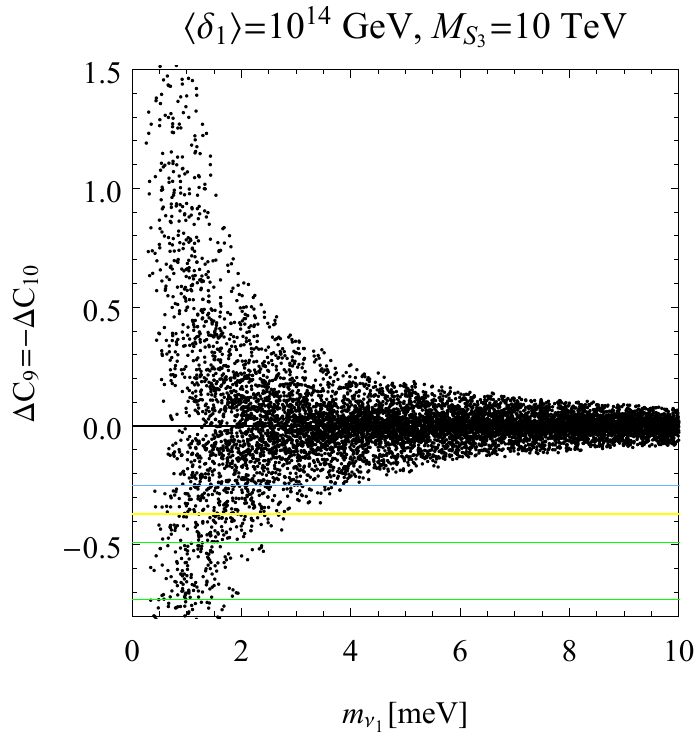}
\caption{
Scan of the $S_3$ contribution to $\Delta C_9 = - \Delta C_{10}$ obtained varying $m_{\nu_1}$ and the PMNS phases $\delta, \alpha, \beta$, as well as the extra phases $\omega_i$ in $V_R$, allowing them to be multiple of $\pi$. The $1\sigma$, $2\sigma$ and $3\sigma$ regions for the explanation of the anomalies are enclosed in green, yellow, and blue lines, respectively.
}
\label{fig:type_I}
\end{figure}

Similarly to the above discussion, we can assume type-I dominance, i.e.~$M_\nu^{II} \ll M_\nu^I$, for example because of a tiny $\langle \delta_3\rangle$. In this case the coupling structure takes the form
\begin{align}
y^R = - \frac{1}{\sqrt{2} \langle \delta_1 \rangle}	 V_R^\intercal m_u U^* \text{diag} \left(\frac{1}{m_{\nu_1}},\frac{1}{m_{\nu_2}},\frac{1}{m_{\nu_3}}\right) U^\dagger m_u V_R \,.
\end{align}
Assuming once again normal ordering, vanishing PMNS phases and tiny $m_{\nu_1}\ll m_{\nu_2}$, as well as $V_R\simeq V_L$, we find
\begin{align}\label{eq:struc_type_I}
y\equiv y^L=y^R \propto \matrixx{0.04 & -0.2 & 4 \\ -0.2 & 1 & -20 \\ 4 & -20 & 380} .
\end{align}
The relative sign between $y^R_{22}$ and $y^R_{23}$ implies a wrong-sign contribution to the neutral-current anomaly (cf.~\eqref{eq:C9}). Taking into account non-zero $m_{\nu_1}$ and PMNS phases $\delta, \alpha, \beta$ does not change this, making it impossible to explain $R_{K^{(*)}}$ using our very restricted model. Taking into account that the general structure of the right-handed CKM-like matrix $V_R$ is given by~\cite{Maiezza:2010ic}
\begin{equation}
V_R \simeq \text{diag}(e^{i \omega_1},e^{i \omega_2},e^{i \omega_3}) \, V_L \, \text{diag}(e^{i \omega_4},e^{i \omega_5},1) \;,
\end{equation}
where the 5 extra phases $\omega_i$ are arbitrary or multiple of $\pi$ according to the discrete symmetry of the model, these phases $\omega_i$ can help in obtaining a correct contribution to $\Delta C_9$, as shown in Fig.~\ref{fig:type_I}.

However, even if one allows for these extra phases and makes use of the various phases to tune $y_{11}$ small in order to suppress $\mu\to e$ conversion, the coupling structure in~\eqref{eq:struc_type_I} does not seem to be successful for the following reason: the 22 and 23 entries, which are the ones that generate $\Delta C_9$, are very suppressed with respect to $y_{33}$. Fitting the anomaly with $M_{S_3} \gtrsim \unit[1]{TeV}$ then requires $y_{33} \gtrsim 3$; even neglecting possible concerns with perturbativity, this large coupling, together with $y_{23}$, would generate a too large contribution to other observables, such $B \to K \nu \nu$ for instance (see also the discussion in the next section). Making use of the various phases does not seem to improve the situation significantly. Therefore, we do not follow this type-I dominance option any further, and instead move to combined explanations of both neutral-current and charged-current anomalies.

We have seen that type-II dominance is compatible with the coupling structure required for $R_{K^{(*)}}$, whereas the type-I contribution is not, at least in the rather stringent PS model with parity considered here. The general case of $M_\nu^{II} \sim M_\nu^I$ introduces more freedom~\cite{Akhmedov:2005np} and will be considered elsewhere in its full extent. We stress again that it is a rather unique feature of PS that allows us to relate LQ couplings to the neutrino mass structure.

\section{Model-building challenges for combined explanations}\label{sec:challenges}
Explaining both sets of anomalies is known to be significantly more challenging than the ones in $b \to s \mu \mu$ alone, essentially because of the low EFT scale $\Lambda \simeq \unit[3.4]{TeV}$ implied by the charged-current anomalies, as compared to $\Lambda \simeq \unit[31]{TeV}$ for the neutral-current ones.

A first possibility would be to try to explain both sets of anomalies by using the vector LQ $U_\mu = (\vec{3},\vec{1})_{2/3}$, present in the adjoint representation of the PS group. One would like to identify this LQ as a PS gauge boson~\cite{Assad:2017iib}. However, this does not work~\cite{DiLuzio:2017vat}, at least in the minimal case, because $U_\mu$, being a gauge boson, would couple with equal strength also to light quarks, and is thus required to be heavier than $\sim \unit[1000]{TeV}$ from the non-observation of new physics in $K_L \to e^{\pm} \mu^{\mp}$, for instance. To circumvent this problem, it has been proposed to generate flavour-dependent couplings to $U_\mu$ by introducing many vector-like fermions, charged under PS, that mix with the SM quarks~\cite{Calibbi:2017qbu} possibly extending the gauge group breaking the lepton-colour unification in $SU(4)_{LC}$~\cite{DiLuzio:2017vat}. A more radical alternative is to consider flavour copies of the PS group (PS$^3$)~\cite{Bordone:2017bld} leading to generation-dependent \emph{gauge} couplings.

Here we attempt a more minimal route by explaining both the anomalies not by the gauge PS leptoquarks, but by the \emph{scalar} ones. Since the coupling of scalar LQ to SM fermions is of Yukawa type, these are naturally flavour dependent without any model-building effort. By limiting oneself to chiral LQs (as suggested by the explanation of $R_K$ vs $R_{K^*}$ assuming new physics in muons~\cite{DAmico:2017mtc}) the only possibilities are the representations in $(\vec{4}, \vec{2}, \vec{1}) \otimes (\vec{4}, \vec{2}, \vec{1})$ plus their chiral partners:
\begin{equation}
(\overline{\vec{10}}, \vec{3}, \vec{1}), \quad (\overline{\vec{10}}, \vec{1}, \vec{1}), \quad ({\vec{6}}, \vec{3}, \vec{1}), \quad ({\vec{6}}, \vec{1}, \vec{1})
\end{equation}
and their chiral partners. The $\vec{6}$ is a real representation of $SU(4)_{LC}$ and as a consequence the $(\vec{\overline{3}},\vec{1})_{1/3}$ within has both leptoquark (from its embedding in $(\vec{6},\vec{1},\vec{1})$) and di-quark couplings (from $(\overline{\vec{6}},\vec{1},\vec{1}) \equiv ({\vec{6}},\vec{1},\vec{1})$), leading to unacceptable fast proton decay. The same holds for the $({\vec{6}},\vec{3},\vec{1})$.

As we saw in the previous section, the $S_3$ in $(\overline{\vec{10}}, \vec{3}, \vec{1})$, which has left-handed couplings, is sufficient to explain the neutral-current anomalies at the scale $\Lambda \simeq \unit[31]{TeV}$. It also contributes to $R_{D^{(*)}}$; however, in order to explain the charged-current anomalies the scale for its contribution to $b \to c \tau \nu$ transitions needs to be $\Lambda \simeq \unit[3.4]{TeV}$. The $SU(2)_L$ conjugate process $b \to s \nu \nu$ would then contribute to $B \to K \nu \nu$, and such a low scale is already excluded by the Belle experiment. Therefore, one could be tempted to also use the $S_1$ in $(\overline{\vec{10}}, \vec{1}, \vec{1})$ with left-handed couplings, analogously to what is done in Refs.~\cite{Crivellin:2017zlb,Buttazzo:2017ixm}, to give a negatively interfering contribution to $B \to K \nu \nu$, bringing this observable in agreement with the existing null results. This does not work in the Pati--Salam model because of the strict constraints posed by the gauge symmetry to the Yukawa couplings. In particular, the couplings of the $(\overline{\vec{10}}, \vec{1}, \vec{1})$ are \emph{antisymmetric} in flavour space. Therefore, in order to explain the charged current anomaly while at the same time suppressing $B \to K \nu \nu$ one needs to use large off-diagonal couplings between second and third generation fermions (and much smaller $y_{33}$). Thus, they would dominantly contribute to $b \to c \tau \nu_\mu$, which is non-interfering with the SM, thus lowering  the scale needed to explain $R_{D^{(*)}}$ to $\Lambda \simeq \unit[0.8]{TeV}$. Hence, the contribution of $S_3$ and $S_1$ must be at similar scales $\approx \unit[]{TeV}$, in order for the cancellation in $B \to K \nu \nu$ to occur, thus leading to an unacceptably large contribution of $S_3$ to the $SU(2)_L$ conjugate transition $b \to s \mu \tau$, and therefore to $B \to K \mu \tau$.

The only possibility left is to use $S_3 \in (\overline{\vec{10}}, \vec{3}, \vec{1})$ and its chiral partner $S_1 \in (\overline{\vec{10}}, \vec{1}, \vec{3})$, which has right-handed non-interfering couplings and therefore cannot be used to follow the strategy above. Nevertheless, in the next section we will show that the anomalies can be explained successfully by assuming that one of the right-handed neutrinos is light enough.

\section{Combined explanation of the \texorpdfstring{$B$}{B}-meson anomalies in Pati--Salam}\label{sec:combined}

As argued in the previous section, we consider the LQ $S_3 \in (\overline{\vec{10}}, \vec{3}, \vec{1})$ with left-handed couplings, and its chiral partner $S_1 \in (\overline{\vec{10}}, \vec{1}, \vec{3})$ with right-handed couplings. Their couplings are given in~\eqref{eq:couplings_S3} and~\eqref{eq:couplings_S1}, and we take them to be real for simplicity. We assume parity, so that $y^L = y^R \equiv y$ and $V_L \simeq V_R$, neglecting the extra phases in $V_R$. 
Dropping the discrete left--right symmetry and considering independent couplings $y_L$ and $y_R$ would obviously make it is easier to accommodate the $B$-meson anomalies, but turns out to not be necessary as a good fit can be obtained even with the strong extra constraint $y^L = y^R$; loop-induced corrections to this relation are discussed in App.~\ref{app:RGE} but do not qualitatively change our conclusions.
We furthermore assume that the couplings of the first generation are negligibly small, $y_{11}\simeq y_{12}\simeq y_{13}\simeq 0$, in order to satisfy the stringent limits involving first-generation fermions discussed in Sec.~\ref{sec:RK}.\footnote{In principle $y_{13}$ can be relatively large, as in Sec.~\ref{sec:RK}, but we set it to zero anyway, to reduce the number of free parameters in the fit, and observables to take into account, since it would basically play no role in the explanation of the anomalies.}

The LQ $S_3$ contributes to $\Delta C_{9} = - \Delta C_{10}$ as given by Eq.~\eqref{eq:C9} and can easily explain the neutral-current anomalies, while $S_1$ does not contribute. To explain also the charged-current anomaly, we need to use $S_1$ with its right-handed couplings and the transition $b_R \to c_R \tau_R N_R$. This contributes to $R_{D^{(*)}}$ if (at least) one of the right-handed neutrino mass eigenstates is lighter than $\sim \unit[100]{MeV}$~\cite{Asadi:2018wea,Greljo:2018ogz,Robinson:2018gza,Azatov:2018kzb}. We will denote this light neutrino mass eigenstate by $\hat N$. Then, assuming that this eigenstate has maximal overlap with the second generation $N_{R,2}$, there is a CKM enhanced contribution to $R_{D^{(*)}}$:
\begin{equation}\label{eq:RD}
R_{D^{(*)}} \ \equiv \ \frac{\Gamma(B \to D \tau \nu,D \tau \hat{N})/\Gamma(B \to D \tau \nu_\tau)_{SM}}{\Gamma(B \to D \hat{\ell} \nu,D \hat{\ell} \hat{N})/\Gamma(B \to D \hat{\ell} \nu_\ell)_{SM}} \, \simeq \, 1 +\left( \frac{v^2}{4 M_{S_1}^2} \frac{V_{R,cs}}{V_{L,cb}} \, y^R_{23}y^R_{23}\right)^2 \stackrel{!}{\simeq} 1.237 \pm 0.053 \,,
\end{equation}
with $\hat{\ell}=e,\mu$. The full expressions for this and the other relevant observables are given in App.~\ref{app:RD}. The relevant transition in flavour space is $b \to c \tau N_2$ and its $SU(2)_R$ conjugate is $b \to s N_3 N_2$. The latter would naively give a too large contribution to $B \to K \nu \nu$, analogously to the discussion in the previous section; however, if we postulate that the mixing between the light eigenstate $\hat N$ and $N_{3}$ is small, a sufficient suppression is obtained. This is the key point of the combined explanation of the $B$-meson anomalies in the PS model with parity. Then, setting  this mixing to zero we have
\begin{equation}\label{eq:BKnunu}
B_{K\nu\nu} \ \equiv \ \frac{\Gamma(B \to K \nu \nu, K \hat{N} \hat{N})}{\Gamma(B \to K \nu_{\ell} \nu_{\ell})_{SM}} \, \simeq \,1+ \left( \frac{2 \pi}{-6.4 \, \alpha \, V_{L,ts} V_{L,tb}} \frac{v^2}{4 M_{S_1}^2} \, y^R_{22} y^R_{23} \right)^2 \, \stackrel{!}{\lesssim} \, 3.3 \,\, (95 \% \,\mathrm{C.L.})
\end{equation}
which can be easily suppressed by $y_{22}$. A small $y_{22} \ll y_{23}$ is also functional to explain the difference in scales of the new-physics effects in $R_{K^{(*)}}$ and $R_{D^{(*)}}$, since the former depends on the flavour combination $y_{22} y_{23}$, as compared to $y_{23} y_{23}$. 

If we allow for a small mixing $(V_R^\dag U_R)_{3 \hat{N}}$, the expression for $R_{D^{(*)}}$ in~\eqref{eq:RD} would be modified by $y_{23}^R \to (V_R^\dag U_R)_{2 \hat{N}} y_{23}^R \,+\, (V_R^\dag U_R)_{3 \hat{N}} y_{33}^R$ with little effect. Instead, in~\eqref{eq:BKnunu} we would have
$y_{22}^R \to (V_R^\dag U_R)_{2 \hat{N}} y_{22}^R \,+\, (V_R^\dag U_R)_{3 \hat{N}} y_{23}^R$, which can potentially increase significantly the new-physics contribution. A simple estimate obtained combining~\eqref{eq:RD} and~\eqref{eq:BKnunu} then gives $(V_R^\dag U_R)_{3 \hat{N}} \lesssim 0.04$, implying a non-generic flavour structure in the right-handed neutrino sector. We will discuss the implications of this in Sec.~\ref{sec:UV}. Notice, however, that this upper limit for the mixing is similar to the CKM one between 2nd and 3rd generations, and therefore this should not be considered a priori as a fine tuning more than the structure of the CKM matrix itself. For simplicity, in the fit we will set $(V_R^\dag U_R)_{3 \hat{N}} = 0$, $(V_R^\dag U_R)_{2 \hat{N}} = 1$ and keep in mind that a non-zero value of the former would shrink the allowed parameter space.

\subsection{Fit of flavour observables}

\begin{table}[t]
\raisebox{2.6em}{\begin{tabular}{c|c}
Parameter & Best fit \\
\hline
$M_{S_3}$ & \unit[6.5]{TeV}\\
$y_{22}$ & 0.034\\
$y_{23}$ & 1.17\\
$y_{33}$ & -0.01\\
\hline
\hline
$\chi^2$ & 2.4\\
\end{tabular}} 
\hspace{10ex}
\begin{tabular}{c|c|c}
Observable & Best fit & Pull/bound \\
\hline
$R_{D^{(*)}}$ & $1.24$ & $+0.0 \sigma$\\
$\Delta C_9 = - \Delta C_{10}$ & $-0.61$ & $+0.0\sigma$\\
$R_{D^{(*)}}^{\mu/e}$ & $1.00$ & $-0.0 \sigma$\\
$\delta g^R_{\tau \tau}$ & $0.5 \times 10^{-4}$ & $-1.2\sigma$\\
$\delta g^R_{\mu \mu}$ & $-9.1 \times 10^{-4}$ & $-0.9\sigma$\\
\hline
$B_{K\nu\nu}$ & $3.25$& $3.28$\\
$\mathrm{BR}(B \to K \mu\tau)$ & $0.9 \times 10^{-5}$ & $4.8 \times 10^{-5}$ \\
$\mathrm{BR}(\tau \to \mu \gamma)$ & $2. \times 10^{-11}$ & $4.4 \times 10^{-8}$\\ 
$\mathrm{BR}(\tau \to 3 \mu)$ & $3. \times 10^{-10}$ & $2.1 \times 10^{-8}$
\end{tabular} 
\caption{Results of the fit for the combined explanation, fixing $M_{S_1} = \unit[1]{TeV}$. \label{tab:fit_combined}}
\end{table}

We perform a combined fit as detailed in App.~\ref{app:RD}. In addition to the anomalies and $B_{K\nu\nu}$, we fit the LFUV in the first two generations as (non) observed in $R_{D^{(*)}}^{\mu/e}$ and the LFV process $B \to K \mu \tau$, which are generated at tree level. Moreover, we include several loop observables: the modifications to the $Z$ couplings $\delta g^R_{\tau \tau}$, $\delta g^R_{\mu \mu}$, and the LFV processes $\tau \to \mu \gamma$, $\tau \to 3 \mu$. One should keep in mind that loop observables can be affected by the other particles in the PS multiplets considered, for instance the other LQs and/or the di-leptons, in case these are light enough. Notice that $\Delta F=2$ processes such as $B_s \leftrightarrow \bar{B}_s$ are small in our setup; this is one of the main virtues of explanations based on scalar LQs.

The results of the fit are given in Table~\ref{tab:fit_combined} for the fixed value of $M_{S_1} = \unit[1]{TeV}$. Clearly, the fit depends only weakly on the absolute mass scale (up to constraints imposed by perturbativity/unitarity), since the tree-level observables are the most constraining and these depend only on ratios of couplings and $M_{S_{1,3}}$. As visible in Table~\ref{tab:fit_combined} the fit is very good and only has some minor pulls from the $Z$ couplings.

\begin{figure}[t]
\includegraphics[scale=1]{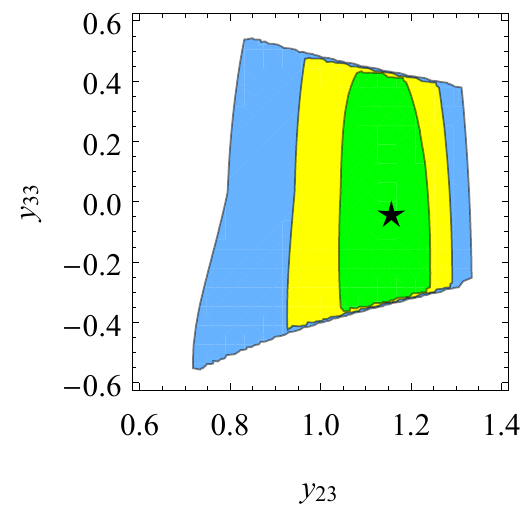}
\hspace{10ex}
\includegraphics[scale=1]{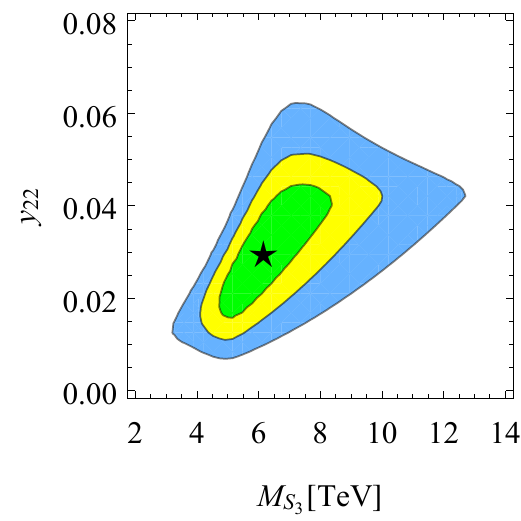}\\[3ex]
\includegraphics[scale=1]{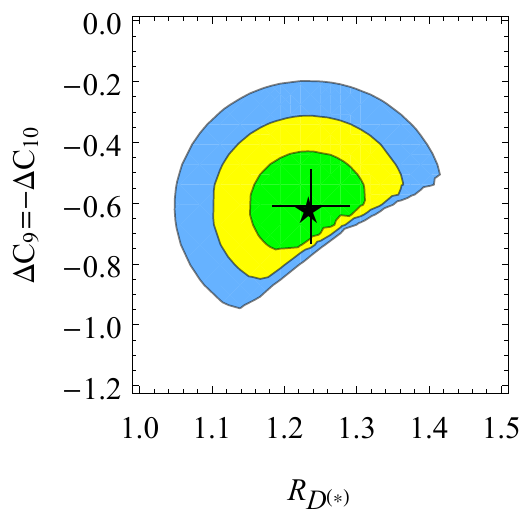}
\hspace{10ex}
\raisebox{1.5ex}{\includegraphics[scale=1.34]{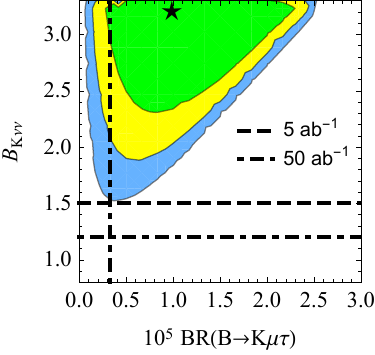}}
\caption{
Results of the fit for the combined explanation, with $M_{S_1} = \unit[1]{TeV}$. We show the $\Delta \chi^2 < 2.3$ ($1 \sigma$), $6.2$ ($2 \sigma$), $11.8$ ($3 \sigma$) regions in green, yellow, and blue, respectively, after marginalising over all other parameters. The star denotes the best-fit point. In the bottom-left panel we show the observed anomalous values for these observables. In the bottom-right one we also show prospects for Belle II for certain luminosities~\cite{BelleIIprospectsLFV,Kou:2018nap}.
}
\label{fig:fit_1_Tev}
\end{figure}

We also show correlations between different parameters and observables in Fig.~\ref{fig:fit_1_Tev}. Note that the charged-current anomaly is fitted basically with zero tension. This is a quite remarkable result since it is often difficult to fit the large observed value even in the much-less restricted simplified models~\cite{Buttazzo:2017ixm}. Note also that one does not need to switch on $y_{33}$, although we did it for the sake of generality. Its inclusion does not improve the quality of the fit significantly. Finally, we stress that one of the phenomenological consequences of the model  is the prediction of a large rate for the LFV process $B \to K \mu \tau$, since the relevant transition $b_L \to s_L \mu_L \tau_L$ is the parity plus $SU(2)_{L}$ partner of the $R_{D^{(*)}}$ one $b_R \to c_R N_{R,2} \tau_R$ (the former is mediated by $S_3$, the latter by $S_1$). In particular,
\begin{equation}\label{eq:BRKmutau}
\mathrm{BR}(B \to K \mu \tau) \ \approx \ 8.6 \times 10^{-3} \, \left(\frac{\unit{TeV}}{M_{S_3}} \right)^4 (y_{23}^L)^4 \ \stackrel{!}{\lesssim} \ 4.8 \times 10^{-5} \, (90 \% \mathrm{C.L.})
\end{equation}
is required to be orders of magnitude higher than in other constructions. This, together with the large rate predicted for $B \to K \nu \nu$ is a smoking-gun phenomenological signature of our framework. As shown in Fig.~\ref{fig:fit_1_Tev} (bottom-right panel), Belle II~\cite{BelleIIprospectsLFV} will test this model completely already with $\unit[5]{ab^{-1}}$ of data (with the assumption of parity).

\subsection{Collider phenomenology}

\begin{figure}[t]
\includegraphics[scale=1.3]{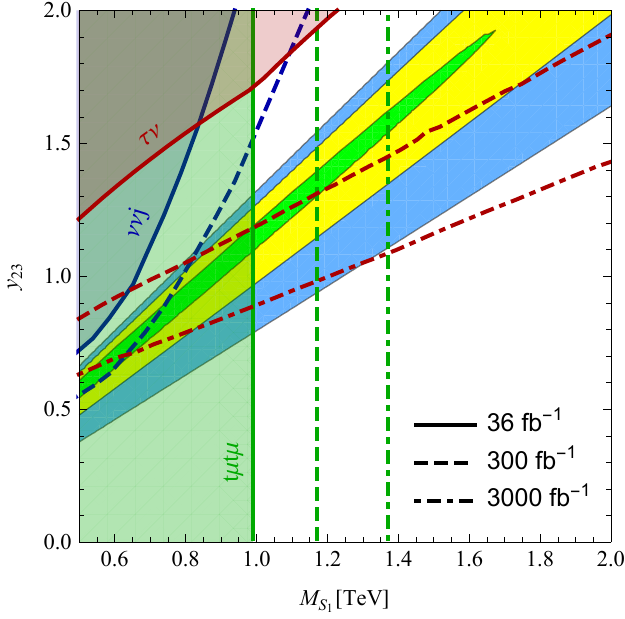}
\caption{
Collider phenomenology for $S_1$. The $1\sigma$, $2\sigma$, $3\sigma$ regions where both sets of anomalies are explained are denoted in green, yellow and blue, respectively. Existing LHC bounds are denoted by continuous lines and shaded regions, future prospects by dotted and dot-dashed lines.
}
\label{fig:collider}
\end{figure}

The LQ $S_1$ needs to be rather light and can be searched for at the LHC. $S_3$ can also be searched for at colliders in the parameter region where it is light enough, but we do not study this possibility here and rather focus on $S_1$. The most important processes are the ones that  involve the SM gauge couplings and/or the LQ Yukawa coupling $y_{23}$, which is required to be large to explain the $B$-meson anomalies. Hence, the most relevant searches are the ones for $pp \to \tau \nu$, $\nu \nu j$ and $t \mu t \mu$. The first two processes are given by a $t$-channel exchange of $S_1$ in the transitions $c \bar{b} \to \tau \hat{N}$, $b \bar{b} \to \hat{N} \hat{N}$, respectively (the latter accompanied by a jet to trigger on it), both proportional to $(y_{23})^2$ for the CKM-unsuppressed channels. The limits for the $\tau \nu$ final state are taken from the ATLAS analysis~\cite{Aaboud:2018vgh}. For the $\nu\nu j$ channel we adapt the results of~\cite{Azatov:2018kzb}, which recasts the CMS monojet search~\cite{Sirunyan:2017jix}. 
The most constraining channel is $pp \to S_1 S_1 \to t \mu t \mu$, where the LQs are pair-produced by QCD interactions. The CMS collaboration has recently performed a dedicated search for LQs decaying into third-generation quarks~\cite{CMS:2018itt}. However, they provide results only for LQs decaying into either $t \mu$ or $b \nu$, i.e.~with $\mathrm{BR}(LQ \to t \mu) + \mathrm{BR}(LQ \to b \nu) = 1$. In our model $S_1$ decays into $t \mu, c \tau, b N$ approximately with the same probability. Therefore, we recast the CMS search in the following way: we extract the results of~\cite{CMS:2018itt} along the line $\mathrm{BR}(LQ \to t \mu) = 1$ and rescale the observed limit on the cross-section by the inverse of our $\mathrm{BR}(S_1 \to t \mu)^2 \simeq 1/9$. We invite the experimental collaborations to analyse more general situations where the total branching ratio is not fixed to 1, and ideally to include the $c \tau$ final state in the analysis. The relation $\mathrm{BR}(S_1 \to t \mu) \simeq \mathrm{BR}(S_1 \to c \tau) \simeq \mathrm{BR}(S_1 \to b \, \text{invisible})$ is a smoking-gun signature of the Pati--Salam model. We show existing bounds and future limits in Fig.~\ref{fig:collider}: the current bound on $M_{S_1}$ is about $\unit[1]{TeV}$. The LHC run III will cover the $1 \sigma$ region, the high-luminosity LHC the entire parameter space that explains the $B$-meson anomalies.

\subsection{Constraints on the UV structure of the model}\label{sec:UV}
The combined explanation of the anomalies poses constraints on the UV structure of the model, in particular regarding the PS breaking and right-handed neutrino mass generation. The successful explanation requires a right-handed neutrino $\hat N$ lighter than $\approx \unit[100]{MeV}$, and therefore much lighter than the PS-breaking scale $\gtrsim \unit[1000]{TeV}$. In the discussion above, where we assumed parity to maximize the predictivity of the model, we also require a non-generic mixing pattern $(V_R^\dag U_R)_{2 \hat{N}} = \mathcal{O}(1)$, $(V_R^\dag U_R)_{3 \hat{N}} \lesssim 0.04$, at the level of the CKM one in the 2nd-3rd generation sector.

First, it is important to stress that this latter requirement is present only if parity is assumed. If the left- and right-handed CKM matrices were largely misaligned, with $V_{R,cb} = \mathcal{O}(1)$, one could explain the charged-current anomaly with an unsuppressed contribution:
\begin{equation}
R_{D^{(*)}} \supset \left( \frac{v^2}{4 M_{S_1}^2} \frac{V_{R,cb}}{V_{L,cb}} \, y^R_{33}y^R_{23}\right)^2 \;,
\end{equation}
which in turns implies a contribution to $B_{K\nu\nu}$ as in~\eqref{eq:BKnunu}, but with Yukawa couplings $y^R_{33} y^R_{23}$ and, most importantly, suppressed by the mixing \emph{squared}  $[(V_R^\dag U_R)_{3 \hat{N}}]^2$. Therefore, in this case the mixing needs to be smaller than $\approx 0.2$, which is not particularly constraining.

The lightness of $\hat{N}$, with $\mathcal{O}(1)$ couplings to $S_1$, requires that $\langle \delta_1 \rangle < \mathcal{O}(100) \unit{MeV}$, if non-zero at all, and therefore this cannot be the only PS-breaking VEV. To show this, let us assume that $\langle \delta_1 \rangle$ is the only source of masses for $N_R$, so that Eq.~\eqref{eq:couplings_S1} gives
\begin{align}
U_R^* \text{diag}(M_1,M_2,M_3) U_R^\dagger = M_R = -\sqrt{2} \langle \delta_1 \rangle V_R y^{R\dagger} V_R^\intercal 
\end{align}
for the right-handed neutrino mass matrix.
The couplings of $S_1$ to the mass eigenstates $N_\text{mass}$ are then
\begin{align}
\frac{1}{2 \, \langle \delta_1 \rangle} \, \overline{d}_R^c [V_R^\intercal U_R \text{diag}(M_1,M_2,M_3) ] N_\text{mass} S_1 \,.
\end{align}
Therefore, in order for the light eigenstate $\hat{N}$ to have $\mathcal{O}(1)$ couplings to $S_1$ its mass cannot be much smaller than $\langle \delta_1 \rangle$. It is not possible to explain both $B$-meson anomalies without assuming additional sources of right-handed neutrino masses, in addition to the VEV $\langle \delta_1 \rangle$, at some high scale.

Fortunately it is not difficult to extend the PS model in order to accommodate light right-handed neutrinos. Most importantly, whatever the mechanism is, its precise form is not expected to affect the explanation of the anomalies given above in this section, as long as it gives a light right-handed neutrino with suppressed mixing to the 3rd generation. For the sake of illustration, we now mention a couple of possibilities, although these should be considered only as toy examples. A detailed study of possible UV sectors goes beyond the scope of this paper.

A first possibility is to simply have a second scalar $(\overline{\vec{10}},\vec{1},\vec{3})$ that gets a high-scale VEV, for instance at the $\unit[10^{12}]{GeV}$ scale, but leaves one of the right-handed neutrinos very light, say $\unit[100]{MeV}$. The light LQs relevant for the $B$-meson anomalies then have Yukawa couplings that are unrelated to the neutrino mass structure and thus allow for $\mathcal{O}(1)$ couplings of this light $N_R$, leading precisely to the scenario fitted above. The large Dirac neutrino masses given by Eq.~\eqref{eq:mass_unification} would then give a too large active neutrino mass of order $m_u^2/(\unit[100]{MeV})$. This type-I contribution could be cancelled by a type-II contribution, which is clearly not well motivated. It is thus necessary to break the relation of Eq.~\eqref{eq:mass_unification} and suppress the Yukawa coupling of the light right-handed neutrino to the active neutrinos, for example by invoking $\phi_{\vec{15}}$. One can then imagine that a Dirac mass $\mathcal{O}(\unit{keV})$ would be suppressed sufficiently by $M_{\hat N} \sim \unit[100]{MeV}$ to generate sub-eV neutrino masses,  giving a mixing angle $\sin \theta = \mathcal{O}(10^{-5})$ between $\hat{N}$ and at least one left-handed neutrino, so that $\hat{N}$ decays into $3 \nu$ in the early universe, with rate~\cite{Adhikari:2016bei}
\begin{equation}
\Gamma(\hat{N} \to 3 \nu) =  \frac{G_F^2  \sin^2 \theta}{96 \pi^3} M_{\hat N}^5 \simeq (\unit[1.5 \times 10^{4}]{sec})^{-1} \left(\frac{\sin \theta}{10^{-5}}\right)^2 \left(\frac{M_{\hat{N}}}{\unit[100]{MeV}}\right)^5\,,
\end{equation}
well after Big-Bang nucleosynthesis,\footnote{Assuming also that it decouples non-relativistically from the thermal plasma, for instance being kept in equilibrium by the process $\hat N \hat N \leftrightarrow \nu_3 \nu_3$, mediated by an $\mathcal{O}(\unit[10]{TeV})$ admixture of the dileptons $\delta_3$ and $\delta_1$.} 
but well before the decoupling of the cosmic microwave background (thus not contributing to dark matter then).
The light $\hat N$ can clearly lead to additional signatures, but generally unrelated to the $B$ anomalies and thus severely model dependent~\cite{Asadi:2018wea,Greljo:2018ogz,Robinson:2018gza,Azatov:2018kzb}. 

A second, more radical, possibility is to get rid of the seesaw mechanism and change the neutrino mass generation. Following Ref.~\cite{Volkas:1995yn} one could break PS by a scalar $(\vec{4},\vec{1},\vec{2})$ and introduce singlet fermions, thus realizing an inverse-seesaw scenario that in principle allows for large couplings of the light $\hat{N}$, and to lower the PS scale down to $\sim \unit[1000]{TeV}$, although in this latter case one needs to introduce also a mechanism to unify left and right gauge couplings at this scale (rather than at $\sim \unit[10^{12}]{TeV}$) in the parity-symmetric scenario. Notice that the issue of neutrino-mass generation (or better, suppression) is often ignored in solutions of the $B$-meson anomalies via the PS-inspired models discussed in Sec.~\ref{sec:challenges}. However, a first attempt in this direction has recently been given in Ref.~\cite{Greljo:2018tuh}, for the model presented in Ref.~\cite{DiLuzio:2017vat}, along the lines discussed in this paragraph.

A full discussion of neutrino masses within PS needs to take the charged-fermion mass generation into account and must include effects of renormalization group running, which in turn depends on all potentially light particles such as the LQs. This is non-trivial even without attempting to resolve the $B$-meson anomalies and will therefore not be investigated here. At least qualitatively it seems that PS could indeed provide a consistent theory behind fermion masses and LFUV.

\section{Conclusion}
\label{sec:conclusion}

Recent hints for lepton-flavour non-universality in various $B$-meson decays point towards new physics around the TeV scale. Combined explanations for $R_{K^{(*)}}$ and $R_{D^{(*)}}$ can be found most easily in leptoquark models, although these are usually ad-hoc extensions without any relations to other processes or structures. Interestingly, they could originate from models that lead to (partial) unification, such as $SU(5)$ or Pati--Salam, where the occurrence of LQs is an absolute necessity.
Here we have shown that the Pati--Salam model naturally contains the relevant scalar LQs to address the $B$-meson anomalies and has additional welcome features: i) proton decay via LQs is avoided, ii) the LQ couplings are automatically chiral, iii) the couplings are restricted and related, making the model highly predictive.

The anomaly in $R_{K^{(*)}}$ can be easily addressed by assuming an $S_3$ LQ around \unit[10]{TeV}. Within the Pati--Salam model the coupling structure is restricted to be \emph{symmetric} in flavour space and can furthermore be related to the type-II seesaw neutrino mass structure, leading to interesting predictions for the neutrino mass parameters. This leads to testable rates for muonic LFV.

Explaining also the anomaly in $R_{D^{(*)}}$ requires one of the right-handed neutrinos to be sub-GeV, as well as an $S_1$ LQ around TeV. Pati--Salam once again heavily restricts the coupling structure, making it non-trivial to explain both $R_{K^{(*)}}$ and $R_{D^{(*)}}$ simultaneously. We have shown that one can obtain a near-perfect explanation for the anomalies despite the restricted couplings, leading again to highly testable predictions for processes such as $B\to K \nu\nu$ and $B\to K \mu\tau$, as well as direct collider searches. The connection to neutrino masses has to be sacrificed; it would be interesting to address it in an extension of the basic Pati--Salam model analyzed here, together with an UV sector of the model that gives rise to the required pattern of masses and mixing at the TeV scale and below.

\section*{Acknowledgements}

We thank Thomas Hambye for discussions and collaboration on the early stages of this project. DT would like to thank Gian Giudice for stimulating discussions that motivated this work, Dario Buttazzo for clarifications on Ref.~\cite{Buttazzo:2017ixm}, and Luca di Luzio for sharing his results on gauge-coupling unification. JH is a postdoctoral researcher of the F.R.S.-FNRS and furthermore supported, in part, by the National Science Foundation under Grant No.~PHY-1620638, and by a Feodor Lynen Research Fellowship of the Alexander von Humboldt Foundation. 
DT is supported by an ULB postdoctoral fellowship, an ULB-ARC grant, the IAP P7/37, and the ERC grant NEO-NAT.

%\vfill\null

\appendix

\section{Details of the fit for \texorpdfstring{$R_{K^{(*)}}$}{R(K)} and type-II neutrino masses}\label{app:RK}
In this appendix we give details about the fit performed in Sec.~\ref{sec:RK} and present the explicit form of the observables used.

For the purposes of the discussion in Sec.~\ref{sec:RK}, we are interested in the contribution of the LQ $S_3$. Its contribution to the neutral-current anomalies is given by~\eqref{eq:C9}. 
The relevant phenomenological constraints are given by observables involving first-generation fermions. In particular, the most stringent constraint is given by the LFV $\mu \to e$ conversion in $\mathrm{Au}_{79}^{197}$ nuclei~\cite{Kitano:2002mt,Dorsner:2016wpm}:
\begin{equation}
\mathrm{BR}(\mu \to e) = \frac{1}{\unit[13.07 \times 10^6]{sec^{-1}}} \frac{4\, M_{\mu}^5}{M_{S_3}^4} \, \bigg| (2 V_p  + V_n) V_{L,u q}^* y^L_{1 q} V_{L,u q'} y^{L*}_{2 q'} + (V_p + 2 V_n) y^L_{11} y^{L *}_{12} \bigg|^2 ,
\end{equation}
with $q,q' = 1,2,3$ being summed over, $V_p = 0.0974$, and $V_n = 0.146$. We also consider the LFV process $\mu \to e \gamma$,
\begin{equation}
\Gamma(\mu \to e \gamma) \ =\ \frac{\alpha \, M_\mu^5}{16384 \pi^4\, M_{S_3}^4} \bigg|-4 y^{L*}_{1 q} y^L_{2 q} \, +\, V_{L,q q'} y^{L*}_{1 q'} V_{L,q q''}^* y^L_{2 q''}  \bigg|^2 ,
\end{equation}
as well as the LFV decay of the neutral kaon $K_L$~\cite{Davidson:1993qk},
\begin{equation}
\Gamma(K_L \to e^\pm \mu^\mp) \ \simeq \ 6.45 \times 10^{-19} \,\unit{GeV} \left(\frac{\unit{TeV}}{M_{S_3}}\right)^4
\, |y^{L*}_{12} y^L_{12} + y^{L*}_{11} y^L_{22}|^2 \;.
\end{equation}
Lepton-flavour \emph{conserving} decays of $K_L$ to light leptons receive an SM contribution that is difficult to evaluate, and therefore we do not consider them. Other potentially large effects are expected in the $R$-ratios measuring LFUV in kaon and pion decays:
\begin{align}
{}^K\!R^{e/\mu} &\equiv \frac{\Gamma(K \to e\nu)/\Gamma(K \to e\nu_e)_{SM}}{\Gamma(K \to \mu\nu)/\Gamma(K \to \mu\nu_\mu)_{SM}} -1  \simeq  - 2 C_3 \, \mathrm{Re} \bigg( \frac{V_{L,uq}}{V_{L,us}} y^{L*}_{1 q} y^L_{12} - \frac{V_{L,uq}}{V_{L,us}} y^{L*}_{2 q} y^L_{22} \bigg) ,\\
{}^\pi\!R^{e/\mu} &\equiv \frac{\Gamma(\pi \to e\nu)/\Gamma(\pi \to e\nu_e)_{SM}}{\Gamma(\pi \to \mu\nu)/\Gamma(\pi \to \mu\nu_\mu)_{SM}} -1  \simeq  - 2 C_3 \, \mathrm{Re} \bigg( \frac{V_{L,uq}}{V_{L,ud}} y^{L*}_{1 q} y^L_{11} - \frac{V_{L,uq}}{V_{L,ud}} y^{L*}_{2 q} y^L_{12} \bigg),
\end{align}
having introduced $C_3 \equiv v^2/(4 M_{S_3}^2)$. Finally, we also consider the contribution to the kaon decay into pions and neutrinos, $K \to \pi \nu \nu$:
\begin{align}
K_{\pi \nu \nu}\ \equiv \ \frac{\Gamma(K^+ \to \pi^+ \nu \nu)}{\Gamma(K^+ \to \pi^+ \nu_{\ell} \nu_{\ell})_{SM}} \ \simeq \ \frac{1}{3} \bigg( &|1 + \,a_K\, C_3 y^L_{12} y^{L*}_{11}|^2 + |1 + \,a_K\, C_3 y^L_{22} y^{L*}_{12}|^2 \notag\\
&+ |1 + \,a_K\, C_3 y^L_{23} y^{L*}_{13}|^2 + |a_K\, C_3 y^L_{22} y^{L*}_{11}|^2 \bigg) ,
\end{align}
with $a_K \equiv \frac{2 \pi}{-6.4 \, \alpha V_{L,td}^* V_{L,ts}}$~\cite{Dorsner:2016wpm}.

To incorporate the information given by experimental \emph{upper bounds} in the fit, we smoothly penalize points in the parameter space that violate a bound by introducing an artificial likelihood
\begin{equation}\label{eq:likelihood}
- 2 \log \mathcal{L}(x= \text{observable}/\text{bound}) =  \left\{ \begin{array}{lr}
0\;, \qquad& x<\lambda\,,\\
\displaystyle 4 \left(\frac{x-\lambda}{1-\lambda}\right)^2 \;, & x \geq \lambda \,,
\end{array}\right.
\end{equation}
with the choice $\lambda = 0.95$, since the real likelihood is not provided by the experiments. In this way, points at the bound are penalized by $- 2 \log \mathcal{L} = 4$, corresponding to $2 \sigma$ in the $\chi^2$ fit, whereas points well within the experimental bound do not receive any penalization. 

\begin{table}
\begin{tabular}{c|c|c}
Observable & Value & Source\\
\hline
$\Delta C_9 = - \Delta C_{10}$ & $-0.61 \pm 0.12$ & \cite{Capdevila:2017bsm}\\
${}^K\!R^{e/\mu}$ & $(4.4 \pm 4.0)\times 10^{-3}$ & \cite{Dorsner:2016wpm,Cirigliano:2007xi}\\
${}^\pi\!R^{e/\mu}$ & $(-2.0 \pm 1.9)\times 10^{-3}$ & \cite{Dorsner:2016wpm,Cirigliano:2007xi}\\
$K_{\pi \nu \nu}$ & $2.2 \pm 1.4$ & \cite{Dorsner:2016wpm,Brod:2010hi} \\
\hline
$\mathrm{BR}(\mu \to e)$ & $< 7 \times 10^{-13}$ & \cite{Bertl:2006up} \\
$\mathrm{BR}(\mu \to e \gamma)$ &  $< 4.2 \times 10^{-13}$ & \cite{TheMEG:2016wtm} \\
$\Gamma(K_L \to e^\pm \mu^\mp)$ & $< \unit[6.0 \times 10^{-29}]{GeV}$ & \cite{Ambrose:1998us} 
\end{tabular}
\caption{Observables used in the fit for $R_{K^{(*)}}$ and neutrino masses in Sec.~\ref{sec:RK}.\label{tab:obs_RK}
}
\end{table}

For the observables, we use the values given in Table~\ref{tab:obs_RK}. In principle, Eq.~\eqref{eq:likelihood} has been chosen to emulate $95\%$ C.L. exclusion limits ($\sim 2 \sigma$), whereas the experimental bounds in Tab.~\ref{tab:obs_RK} are at $90\%$ C.L.; we ignore this difference because the experimental bounds at these different C.L. are presumably very close to each other for these observables.

\section{Details of the fit for \texorpdfstring{$R_{K^{(*)}}$}{R(K)} and \texorpdfstring{$R_{D^{(*)}}$}{R(D)}}\label{app:RD}

In this appendix we give details about the fit performed in Sec.~\ref{sec:combined} and give the explicit form of the observables used.
We are interested in the contribution of the LQs $S_3$ and $S_1$ with left- and right-handed couplings, respectively. The couplings are taken to be real, and the ones to the first generation are considered to be negligible.

Only $S_3$ contributes to the neutral-current anomaly, as given by Eq.~\eqref{eq:C9}. Instead, both of them contribute to the charged-current anomaly:
\begin{equation}
R_{D^{(*)}} \ \equiv \ \frac{\Gamma(B \to D \tau \nu,D \tau \hat{N})/\Gamma(B \to D \tau \nu_\tau)_{SM}}{\Gamma(B \to D \hat{\ell} \nu,D \hat{\ell} \hat{N})/\Gamma(B \to D \ell \nu_\ell)_{SM}} \, = \frac{R_{D^{(*)}}^\tau}{\frac{1}{2}(1+R_{D^{(*)}}^{\mu})}
\end{equation}
with 
\begin{align}
R_{D^{(*)}}^\tau &= \left( 1 - C_{3} y^L_{33} \left(\frac{V_{L,cs}}{V_{L,cb} }y^L_{23} + y^L_{33} \right)\right)^2 \, +\, \left(C_{3} y^L_{23} \left(\frac{V_{L,cs}}{V_{L,cb} }y^L_{23} + y^L_{33} \right)\right)^2 \notag\\
&\quad+ \,  \left(C_{1} y^R_{23} \left( \frac{V_{R,cs}}{V_{L,cb}} y^R_{23} + \frac{V_{R,cb}}{V_{L,cb}} y^R_{33}  \right) \right)^2 ,\\
R_{D^{(*)}}^\mu &= \left( 1 - C_{3} y^L_{23} \left(\frac{V_{L,cs}}{V_{L,cb} }y^L_{22} + y^L_{23} \right)\right)^2 \, +\, \left(C_{3} y^L_{33} \left(\frac{V_{L,cs}}{V_{L,cb} }y^L_{22} + y^L_{23} \right)\right)^2 \notag\\
&\quad+ \,  \left(C_{1} y^R_{23} \left( \frac{V_{R,cs}}{V_{L,cb}} y^R_{22} + \frac{V_{R,cb}}{V_{L,cb}} y^R_{23}  \right) \right)^2 ,
\end{align}
and $C_{1,3} \equiv v^2/(4 M_{S_{1,3}}^2)$. The contribution of $S_1$ dominates, since it is significantly lighter than $S_3$.
We also consider the LFUV in the first two generations:
\begin{equation}
R_{D^{(*)}}^{\mu/e} \ \equiv \ \frac{\Gamma(B \to D \mu \nu,D \mu \hat{N})/\Gamma(B \to D \mu \nu_\mu)_{SM}}{\Gamma(B \to D e \nu,D e \hat{N})/\Gamma(B \to D e \nu_e)_{SM}} \, = R_{D^{(*)}}^\mu \,.
\end{equation}
As discussed in the main body, a particularly constraining observable is given by the decay of $B$ into a kaon and neutrinos
\begin{align}
B_{K\nu \nu}\ \equiv \ \frac{\Gamma(B \to K \nu \nu, K \hat{N} \hat{N})}{\Gamma(B \to K \nu_{\ell} \nu_{\ell})_{SM}} \ &= \ \frac{1}{3} \bigg[1 + (1 + \,a\, C_3 y^L_{33} y^L_{23})^2 + (1 + a \,C_3\, y^L_{23} y^L_{22})^2\notag\\
&\quad  + (a \,C_3\, y^L_{33} y^L_{22})^2 + (a \,C_3\, y^L_{23} y^L_{23})^2 + (a \, C_1\, y^R_{22} y^R_{23})^2 \bigg] ,
\end{align}
with $a \equiv \frac{2 \pi}{-6.4 \, \alpha V_{L,ts} V_{L,tb}}$. Since the  coupling $y_{23}$ needs to be relatively large to explain the anomalies, we have an enhanced contribution to the LFV decay~\cite{Dorsner:2017ufx}:
\begin{align}
\mathrm{BR}(B \to K \mu \tau) \simeq 8.6 \times 10^{-3} \, \left( \frac{\unit{TeV}}{M_{S_3}}\right)^4 \, \left[(y^L_{23} y^L_{23})^2 + (y^L_{22} y^L_{33})^2 \right] .
\end{align}
We also consider loop modifications of the $Z$-boson couplings to right-handed leptons, normalized as $\tfrac{g }{c_W} (s_W^2+\delta g^R_{\ell \ell'}) Z_\alpha \overline{\ell} \gamma^\alpha P_R \ell'$:
\begin{align}
\delta g^R_{\tau \tau} &\simeq \frac{1}{288 \pi^2   M_{S_1}^2} \bigg[ 27 M_t^2 (V_{R,ts} y^R_{23} + V_{R,tb} y^R_{33})^2 \left(1 - 2 \, \log\left( \frac{M_{S_1}}{M_t}\right) \right) \notag\\
&\quad +  M_Z^2 s_W^2 \Big( (V_{R,cs} y^R_{23} + V_{R,cb} y^R_{33})^2 + (V_{R,ts} y^R_{23} + V_{R,tb} y^R_{33})^2\Big) \left(5 + 24 \, \log\left( \frac{M_{S_1}}{M_Z}\right) \right)\bigg] ,\\
\delta g^R_{\mu \mu} &\simeq \frac{1}{288 \pi^2  M_{S_1}^2} \bigg[ 27 M_t^2 (V_{R,ts} y^R_{22} + V_{R,tb} y^R_{23})^2 \left(1 - 2 \, \log\left( \frac{M_{S_1}}{M_t}\right) \right) \notag\\
&\quad +  M_Z^2 s_W^2 \Big( (V_{R,cs} y^R_{22} + V_{R,cb} y^R_{23})^2 + (V_{R,ts} y^R_{22} + V_{R,tb} y^R_{23})^2\Big) \left(5 + 24 \, \log\left( \frac{M_{S_1}}{M_Z}\right) \right)\bigg] ,\\
\delta g^R_{\mu \tau} &\simeq \frac{1}{576 \pi^2  M_{S_1}^2} \bigg[ 3 (V_{R,ts} y^R_{22} + V_{R,tb} y^R_{23}) (V_{R,ts} y^R_{23} + V_{R,tb} y^R_{33}) \notag\\
&\quad\times \bigg[18 M_t^2 - M_Z^2 (3 c_W^2+13s_W^2) - 4 (9 M_t^2 - 4M_Z^2 s_W^2) \log \left( \frac{M_{S_1}}{M_t} \right)\bigg] \notag\\
&\quad + 2 M_Z^2 s_W^2 (V_{R,cs} y^R_{22} + V_{R,cb} y^R_{23})(V_{R,cs} y^R_{23} + V_{R,cb} y^R_{33})\left(5 + 24 \, \log\left( \frac{M_{S_1}}{M_Z}\right) + 12  \, i\, \pi \right)\bigg] ,
\end{align}
having neglected the non-interfering imaginary part for the lepton-flavour conserving couplings.
Similar corrections but to the left-handed couplings $\delta g^L_{\ell \ell'}$ arise from $S_3$ but are correspondingly suppressed. The LFV coupling $\delta g^R_{\mu \tau}$ gives rise to the LFV decay of the $\tau$ lepton,
\begin{equation}
\frac{\Gamma(\tau \to \mu \mu \mu)}{\Gamma(\tau \to \mu \nu \nu)} \simeq \left[8 s_W^4 + 4 (s_W^2 - 1/2)^2 \right] \, |\delta g^R_{\tau \mu}|^2 ,
\end{equation}
neglecting additional box diagrams~\cite{Gabrielli:2000te}.
Finally, we also consider the radiative LFV decay $\tau \to \mu \gamma$,
\begin{align}
\Gamma(\tau \to \mu \gamma) &\simeq \frac{\alpha \, (M_\tau^2 - M_\mu^2)^3}{16384\,\pi^4 M_{S_1}^4 M_\tau} \, \big[(V_{R,ts} y^R_{22} + V_{R,tb} y^R_{23}) \, (V_{R,ts} y^R_{23} + V_{R,tb} y^R_{33}) \notag\\
&\quad+ (V_{R,cs} y^R_{22} + V_{R,cb} y^R_{23}) \, (V_{R,cs} y^R_{23} + V_{R,cb} y^R_{33}) \big]^2 ,
\end{align}
again neglecting the contribution of $S_3$.

\begin{table}
\begin{tabular}{c|c|c}
Observable & Value & Source\\
\hline
$\Delta C_9 = - \Delta C_{10}$ & $-0.61 \pm 0.12$ & \cite{Capdevila:2017bsm}\\
$R_{D^{(*)}}$ & $1.237 \pm 0.053$ & \cite{Buttazzo:2017ixm} \\
$R_{D^{(*)}}^{\mu/e}$ & $1.00 \pm 0.02$ & \cite{pdg} \\
$\delta g^R_{\tau \tau}$ & $(8 \pm 6) \times 10^{-4}$& \cite{ALEPH:2005ab}\\
$\delta g^R_{\mu \mu}$ & $(3 \pm 13) \times 10^{-4}$& \cite{ALEPH:2005ab}\\
\hline
$B_{K\nu\nu}$ & $<3.28$ & \cite{Grygier:2017tzo}\\
$\mathrm{BR}(B \to K \mu \tau)$ & $< 4.8 \times 10^{-5}$ & \cite{Lees:2012zz}\\
$\mathrm{BR}(\tau \to \mu \gamma)$ & $< 4.4 \times 10^{-8}$ & \cite{Aubert:2009ag}\\
$\mathrm{BR}(\tau \to \mu \mu \mu)$ & $< 2.1 \times 10^{-8}$ & \cite{Hayasaka:2010np}\\
\end{tabular}
\caption{Observables used in the combined fit in Sec.~\ref{sec:combined}. $B_{K\nu\nu}$ was converted to $95\%$ C.L., see the discussion in the text, all other upper bound are $90\%$ C.L. \label{tab:obs_combined}}
\end{table}

For the observables, we use the values given in Table~\ref{tab:obs_combined}. We incorporate the information given by experimental upper bounds in the fit by means of the artificial likelihood~\eqref{eq:likelihood}. Since this is meant to emulate $95\%$ C.L. exclusion limits ($\sim 2 \sigma$), we chose to convert naively the $90 \%$C.L. bound for $B_{K\nu\nu}$ given by Belle~\cite{Grygier:2017tzo}, as $2.7 \times (2\sigma/1.645\sigma) \simeq 3.28$, in order to be conservative for this particularly constraining observable. We ignore this difference for the other observables.

\section{Renormalization group running of Yukawas}
\label{app:RGE}

The most interesting LQ-coupling predictions of our Pati--Salam model are the symmetric structure $y=y^{\intercal}$ as well as the left--right exchange symmetry $y^L=y^R$ (although this can be loosened even within PS). These relations hold above the PS-breaking scale $\mu_\text{PS}$ by construction, but will typically be broken at lower scales due to loop effects. In this appendix we will study this in two benchmark scenarios. For this purpose we make use of the one-loop renormalization group equations (RGEs) for the LQ Yukawa couplings $y$, $\dd y/\dd \log \mu = \beta_y/(4\pi)^2$, which have been known for a long time~\cite{Cheng:1973nv,Machacek:1983fi} and have been obtained here with SARAH~\cite{Staub:2013tta,Staub:2015kfa}. We will make a number of simplifying assumptions in order to focus on the main points: we will assume at low scales just the SM plus the $S_x$ LQs and in some cases the right-handed neutrinos. All other particles, namely the non-SM gauge bosons and scalars, are assumed to sit closer to the PS scale and are thus neglected in the running. For the sake of simplicity, we assume that the SM Higgs boson $H$ contributes predominantly to the SM fermion masses, thus approximating the Yukawa couplings by $y_f = m_f/\langle H\rangle$. In this approximation, only the top-quark Yukawa is large and relevant for the RGEs, so we will neglect all the other SM Yukawa couplings.
Note that the LQ Yukawa matrices are written so that they have the quark index on the left side and the lepton index on the right, as is clear from the matrix multiplications below.

\subsection{Light right-handed neutrinos}

Let us start our discussion by assuming the right-handed neutrinos to be light enough to contribute to the RGEs down to the TeV scale. This is motivated by the fact that at least one of the right-handed neutrinos needs to be light for the combined explanation of the anomalies, as described in Sec.~\ref{sec:combined}.
The beta functions for all the LQ couplings are then
\begin{align}
%%%%%%%%
\beta_{y^L_{S_3}} &= -\left( \frac{1}{2} g_1^2+\frac{9}{2} g_2^2 +4 g_3^2 \right) y^L_{S_3} +\tr \left(y^L_{S_3} y^{L\dagger}_{S_3} \right) y^L_{S_3}  + 3 y^L_{S_3} y^{L\dagger}_{S_3} y^L_{S_3}  + \frac{1}{2} y_u^\intercal y_u^* y^L_{S_3} \,,\\
%%%%%%%%%
\begin{split}
\beta_{y^{R}_{\tilde S_1}} &= -\left( 2 g_1^2 +4 g_3^2 \right) y^{R}_{\tilde S_1} +\tr \left( y^{R}_{\tilde S_1} y^{R \dagger}_{\tilde S_1} \right) y^{R}_{\tilde S_1}  + 2 y^{R}_{\tilde S_1} y^{R \dagger}_{\tilde S_1}  y^{R}_{\tilde S_1}  + \frac{3}{4} y^{R}_{\tilde S_1} y^{R\dagger}_{S_{1,e}} y^R_{S_{1,e}} \\
&\quad+\frac{1}{4} y^R_{S_{1,N}} y^{R\dagger}_{S_{1,N}} y^{R}_{\tilde S_1}  \,,
\end{split}\\
%%%%%%%%
\begin{split}
\beta_{y^{R}_{S_{1,e}}} &= -\left( \frac{13}{5} g_1^2 +4 g_3^2 \right) y^{R}_{S_{1,e}} +\frac{1}{2} \tr \left( y^{R}_{S_{1,e}} y^{R\dagger}_{S_{1,e}} \right) y^{R}_{S_{1,e}} +\frac{1}{2}\tr \left(  y^{R}_{S_{1,N}} y^{R\dagger}_{S_{1,N}}\right)  y^{R}_{S_{1,e}}\\
&\quad + y^{R}_{S_{1,e}} y^{R\dagger}_{S_{1,e}}  y^{R}_{S_{1,e}} +\frac{3}{2} y^{R}_{S_{1,e}} y^{R \dagger}_{\tilde S_1} y^{R}_{\tilde S_1}+\frac{1}{2} y^{R}_{\bar S_1} y^{R\dagger}_{\bar S_1}  y^{R}_{S_{1,e}} + y_u y_u^\dagger y^{R}_{S_{1,e}} \,,
\end{split}\\
%%%%%%%%
\begin{split}
\beta_{y^{R}_{S_{1,N}}} &= -\left( \frac{1}{5} g_1^2 +4 g_3^2 \right) y^{R}_{S_{1,N}} +\frac{1}{2} \tr \left( y^{R}_{S_{1,N}} y^{R\dagger}_{S_{1,N}} \right) y^{R}_{S_{1,N}} +\frac{1}{2}\tr \left(  y^{R}_{S_{1,e}} y^{R\dagger}_{S_{1,e}}\right)  y^{R}_{S_{1,N}}\\
&\quad + y^{R}_{S_{1,N}} y^{R\dagger}_{S_{1,N}}  y^{R}_{S_{1,N}} +\frac{3}{2} y^{R}_{S_{1,N}} y^{R\dagger}_{\bar S_1} y^{R}_{\bar S_1}+\frac{1}{2} y^{R}_{\tilde S_1} y^{R \dagger}_{\tilde S_1}  y^{R}_{S_{1,N}} \,,
\end{split}\\
%%%%%%%%
\begin{split}
\beta_{y^{R}_{\bar S_1}} &= -\left( \frac{4}{5} g_1^2 +4 g_3^2 \right) y^{R}_{\bar S_1} + \tr \left( y^{R}_{\bar S_1} y^{R\dagger}_{\bar S_1} \right) y^{R}_{\bar S_1}+ 2 y^{R}_{\bar S_1} y^{R\dagger}_{\bar S_1}  y^{R}_{\bar S_1} +\frac{3}{4} y^{R}_{\bar S_1} y^{R\dagger}_{S_{1,N}} y^{R}_{S_{1,N}} \\
&\quad +\frac{1}{4} y^{R}_{S_{1,e}} y^{R\dagger}_{S_{1,e}}  y^{R}_{\bar S_1} + y_u y_u^\dagger y^{R}_{\bar S_1} \,,
\end{split}
\end{align}
where we introduced two different couplings for $S_1$,  which are identical above the PS scale but split by the RG flow below it. All these couplings should be equal at the scale $\mu_\text{PS}$, which serves as a boundary condition for the solution of these differential equations. We will be interested in the coupling structures at scales $\mu \sim \unit{TeV}$, where the anomalies are generated. At this low scale, one of the coupling matrices can be fixed, e.g.~by our fit, but the other ones are determined by RGE running.

\begin{figure}[t]
\includegraphics[height=15em]{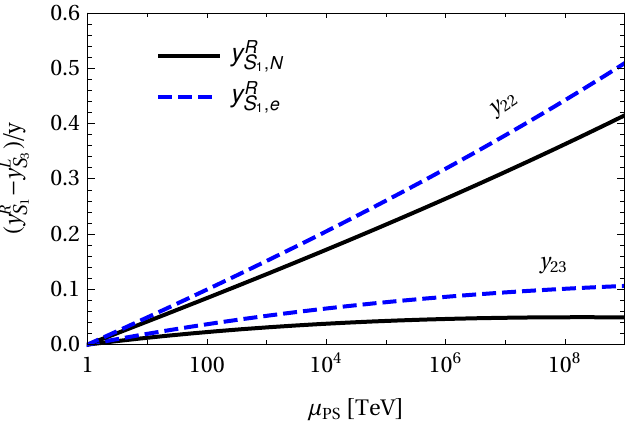}
\hspace{2ex}
\includegraphics[height=15em]{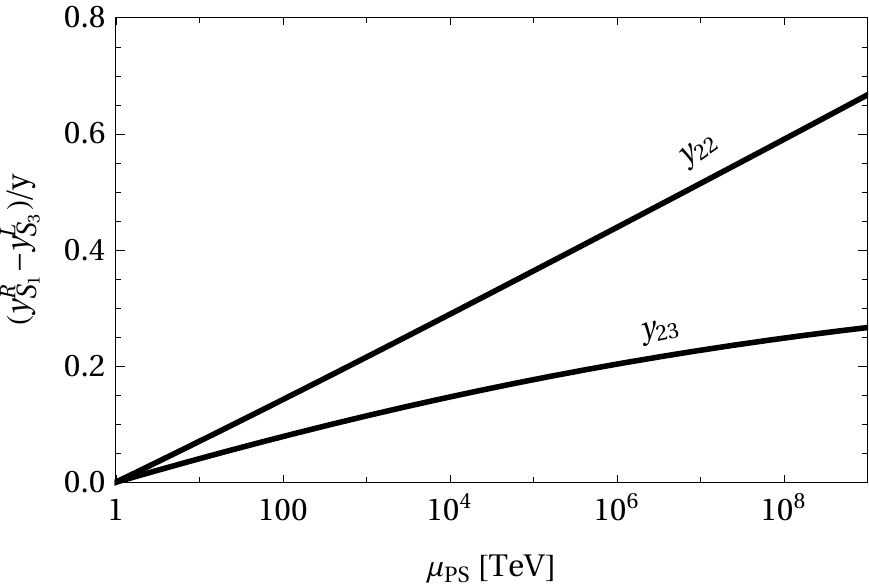}
\caption{
The relative difference of $S_3$ and $S_1$ LQ couplings at the TeV scale, $(y^R - y^L)/y^L|_{\mu =\unit{TeV}}$, as a function of the PS scale $\mu_\text{PS}$. The left (right) figure assumes light (heavy) right-handed neutrinos.}
\label{fig:running}
\end{figure}

All couplings receive the same contribution from the strong-coupling constant $g_3$, but will clearly run apart due to the different contributions from the electroweak gauge couplings $g_{1,2}$ and the top-quark Yukawa $y_u = \diag (0,0,y_\text{top})$. (Note that $g_1$ has the standard GUT normalization.)
In the limit $g_{1,2}=0=y_u$, $y^R_{\dots}=y^R$  all the $y^R$ beta functions become identical and equal to the one for $y^L$. This is to be expected because $(S_1,\bar S_1,\tilde S_1)$ forms the parity partner of $S_3$.
The PS predictions $y^R_{\dots} = y^L = (y^{L})^\intercal$ are then mainly broken by $y_\text{top}\sim 1$ with contributions of order $y \frac{y_\text{top}^2}{(4\pi)^2} \log (\mu_\text{PS}/\unit{TeV})$. This corresponds to percent-level modifications of our formula even if the PS scale is close to the GUT scale.

\begin{figure}[t]
\includegraphics[height=15em]{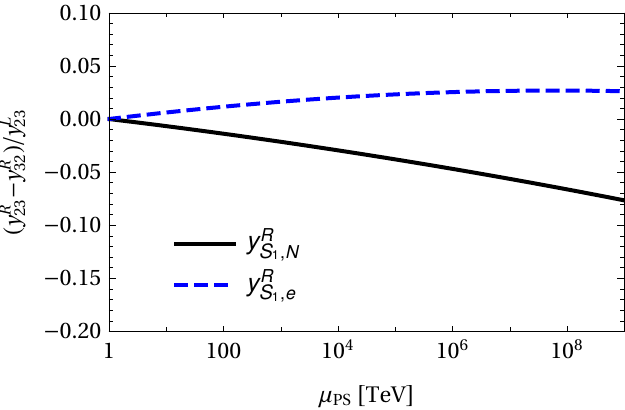}
\caption{
The relative difference of $23$ and $32$ LQ couplings at the TeV scale, $(y^R_{23} - y^R_{32})/y^L_{23}|_{\mu =\unit{TeV}}$, as a function of the PS scale $\mu_\text{PS}$, assuming light right-handed neutrinos.}
\label{fig:running_asym}
\end{figure}

This conclusion makes use of the fact that light right-handed neutrinos as well as light $S_3$ and $(S_1,\bar S_1,\tilde S_1)$ LQs leave the left--right exchange symmetry approximately intact. Let us consider an explicit breaking of this symmetry by making $\bar S_1$ and $\tilde S_1$ heavy, leaving only $S_3$, $S_1$, and the light $N$s in addition to the SM. This case, corresponding to $y^{R}_{\bar S_1} =0=y^{R}_{\tilde S_1}$ in the above RGEs, is shown in Fig.~\ref{fig:running} (left), assuming $y^L (\mu=\unit{TeV})$ to be of the form relevant for our fit, i.e.~taking the best-fit values from Tab.~\ref{tab:fit_combined}. We see that the absence of $\bar S_1$ and $\tilde S_1$ enhances the left--right asymmetry $y^R-y^L$, albeit only at the level of $\mathcal{O}(10\%)$ for not too large PS scales. Such corrections are not relevant for the fit performed in the main text and do not change our conclusions. This is especially true for the 23 entry, which is the largest of the Yukawa couplings and is left--right symmetric at the few percent level. Finally, in Fig.~\ref{fig:running_asym} we show the relative breaking of the most authentic PS prediction $y=y^\intercal$, i.e.~colour-lepton unification, due to the running; again, for the 23 vs 32 entries, relevant for the explanation of the anomalies, the asymmetry is at most at the few percent level, as anticipated above.

\subsection{Heavy right-handed neutrinos}

Let us now consider the case in which all right-handed neutrinos are heavy. While, as pointed out above, at least one of the right-handed neutrinos needs to be light for the combined explanation of the anomalies, the case considered here provides a benchmark scenario to assess the dependence of our conclusions on the details of the spectrum not relevant for the explanation of the anomalies. 
The heaviness of the right--handed neutrinos leads to a larger violation of the PS predictions $y^R_{\dots} = y^L$. Assuming only the SM plus the $S_x$ LQs we find the RGEs
\begin{align}
\beta_{y^L_{S_3}} &= -\left( \frac{1}{2} g_1^2+\frac{9}{2} g_2^2 +4 g_3^2 \right) y^L_{S_3} +\tr \left(y^L_{S_3} y^{L\dagger}_{S_3} \right) y^L_{S_3}  + 3 y^L_{S_3} y^{L\dagger}_{S_3} y^L_{S_3}  + \frac{1}{2} y_u^\intercal y_u^* y^L_{S_3} \,,\\
%%%%%%%%
\beta_{y^{R}_{\tilde S_1}} &= -\left( 2 g_1^2 +4 g_3^2 \right) y^{R}_{\tilde S_1} +\tr \left( y^{R}_{\tilde S_1} y^{R \dagger}_{\tilde S_1} \right) y^{R}_{\tilde S_1}  + 2 y^{R}_{\tilde S_1} y^{R \dagger}_{\tilde S_1}  y^{R}_{\tilde S_1}  +\frac{3}{4} y^{R}_{\tilde S_1} y^{R\dagger}_{S_1} y^R_{S_1}  \,,\\
%%%%%%%%
\beta_{y^{R}_{S_1}} &= -\left( \frac{13}{5} g_1^2 +4 g_3^2 \right) y^{R}_{S_1} +\frac{1}{2} \tr \left( y^{R}_{S_1} y^{R\dagger}_{S_1} \right) y^{R}_{S_1} + y^{R}_{S_1} y^{R\dagger}_{S_1}  y^{R}_{S_1}  +\frac{3}{2} y^{R}_{S_1} y^{R \dagger}_{\tilde S_1} y^{R}_{\tilde S_1}  + y_u y_u^\dagger y^{R}_{S_1} \,,
\end{align}
with $\bar S_1$ being irrelevant since it only has couplings to the heavy $N$. We will further assume $\tilde S_1$ to be heavy, since we do not need it for the $B$-meson anomalies, leaving only the $S_1$ and $S_3$ Yukawa couplings to be compared. These two couplings will again run apart, more prominently than before due to all the additional explicit left--right breaking due to split multiplets (see Fig.~\ref{fig:running} (right)). The large 23 entry is still roughly the same for $S_1$ and $S_3$, the small $\mathcal{O}(20\%)$ corrections not being important in our fit. 
The 22 entry becomes potentially more asymmetric, but still at most at the $\mathcal{O}(60\%)$ level. Therefore, even in the most pessimistic case, our results in Fig.~\ref{fig:fit_1_Tev} will be at most changed by $\mathcal{O}(1)$ factors, leaving our qualitative conclusions  unchanged.

Finally, we conclude by stressing the fact that the above exercise should be considered as a simple estimate of the accuracy of the approximations used in the main body. A quantitatively accurate treatment of the running requires the specification of the full spectrum up to the PS scale. However, only particles lighter than a few TeV contribute to the anomalies and therefore we do not need to specify the full spectrum for the purposes of this work. This will be left for future work, but below we will provide a brief discussion on the impact of $y_L\neq y_R$ on the $B$-anomaly fit.

\subsection{Impact on the fit for the combined explanations}

\begin{figure}[t]
\includegraphics[scale=1]{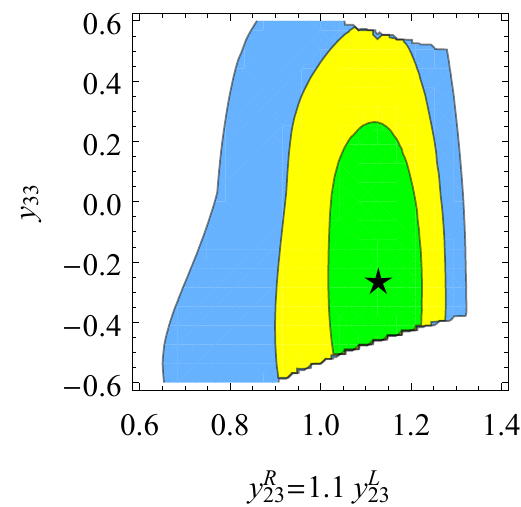}
\hspace{10ex}
\includegraphics[scale=1]{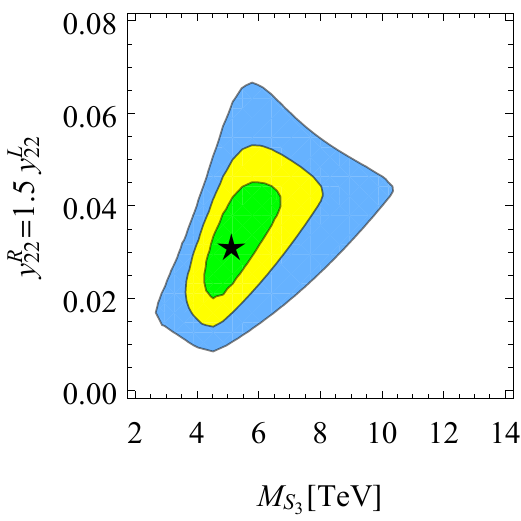}\\[3ex]
\includegraphics[scale=1]{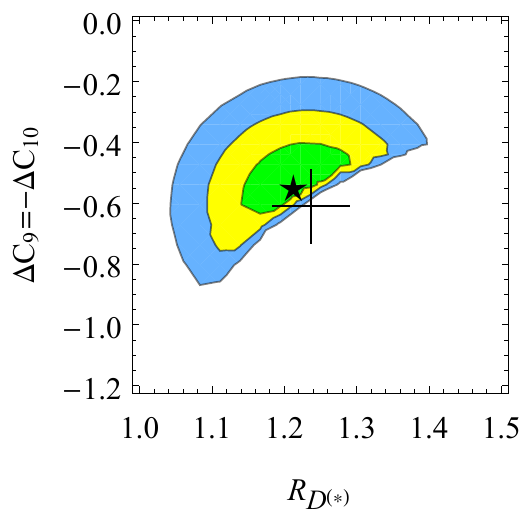}
\hspace{10ex}
\raisebox{1.5ex}{\includegraphics[scale=1.34]{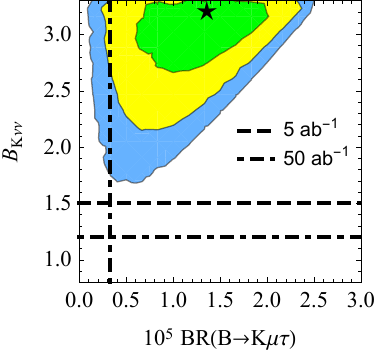}}
\caption{
Same as Fig.~\ref{fig:fit_1_Tev}, but with the couplings deformed as in Eq.~\eqref{eq:deform}.
}
\label{fig:fit_1_Tev_bis}
\end{figure}

As clear from the discussion above, the relations $y^R_{\dots} = y^L = (y^{L})^\intercal$ are deformed at the TeV scale from RG effects. Nevertheless, in our assumption of a left--right symmetry at the PS scale, the two sets of couplings should not be considered independent for the combined fit as in Sec.~\ref{sec:combined}. 

Knowing the full spectrum up to the PS scale one could then consider the couplings $y^R = y^L = (y^{L})^\intercal$ at the PS scale as input parameters and, for each point in the parameter space, perform the RG running down to the TeV scale and use these value for the phenomenological observables entering the fitting procedure. However, since we do not presume to have knowledge of the full spectrum (only the LQs at the TeV scale have direct impact on the explanation of the anomalies), such a procedure would hardly seem sensible considering the combinatorial number of possible light states and thresholds. Moreover, from a more practical point of view, performing the RG running for each point in the parameter space during the likelihood maximization procedure would increase significantly the computing power required for the numerical fit.

Therefore, in order to assess the impact of the RG-running effects, we perform a fit with the relations $y^R_{\dots} = y^L = (y^{L})^\intercal$ broken by a  fixed typical amount. In detail, as shown in Fig.~\ref{fig:running} and~\ref{fig:running_asym} assuming light right-handed neutrinos, the main breaking is in the equality $y^R = y^L$ for the relevant 22 and 23 entries. We thus choose the deformation suggested by the running of the best-fit values from Tab.~\ref{tab:fit_combined} (see Fig.~\ref{fig:running} (left))
\begin{align}\label{eq:deform}
y^R_{22} = 1.5 \times y^L_{22} \; , \quad y^R_{23} = 1.1 \times y^L_{23} \,,
\end{align}
with equal couplings of $S_1$ to $N$ and $e$ and keeping the rather accurate PS relation $y=y^\intercal$. The results of the fit are shown in Fig.~\ref{fig:fit_1_Tev_bis}. As can be seen by comparing with Fig.~\ref{fig:fit_1_Tev}, the impact of RG effects is rather mild, thus justifying the usage of the unbroken relations $y^R_{\dots} = y^L = (y^{L})^\intercal$ for qualitative and semi-quantitative conclusions, as in the main body of the paper.

\section{Left--right gauge coupling unification}
\label{app:RG_gauge}

The PS group breaking $SU(4)_{LC} \times SU(2)_L \times SU(2)_R\to SU(3)_C\times SU(2)_L \times U(1)_Y$ by itself does not impose any restrictions on the three gauge couplings. However, since we mostly focus on the PS model \emph{with} left--right unification (parity $L\leftrightarrow R$) in order to constrain the left and right Yukawa couplings for a \emph{combined} explanation of both sets of anomalies in Sec.~\ref{sec:combined}, one may wonder if such a scenario is viable from the UV point of view. In particular, this requires the gauge couplings of the $SU(2)_L$ and $SU(2)_R$ factors to be equal at a scale that defines the PS-breaking scale $\mu_\text{PS}$,  $g_R(\mu_\text{PS})=g_L(\mu_\text{PS})$, which translates into a matching condition on the SM gauge-coupling fine-structure constants $\alpha_i \equiv g_i^2/(4 \pi)$,
\begin{align}\label{eq:unification}
\frac{3}{5} \, \alpha_L^{-1}(\mu_\text{PS})  =  \alpha_{1}^{-1}(\mu_\text{PS}) - \frac{2}{5} \alpha_{C}^{-1}(\mu_\text{PS}) \,,
\end{align}
with GUT normalization for the hypercharge generator $Q_1 = \sqrt{3/5} \, Y$. This is the lowest-order matching condition appropriate for the one-loop RGEs under consideration here, neglecting threshold corrections, see e.g.~Refs.~\cite{Weinberg:1980wa,Hall:1980kf,Bertolini:2009qj} for improved formulae. Using only the SM particle content this fixes the PS scale $\mu_\text{PS}\sim \unit[5\times 10^{13}]{GeV}$.

\begin{figure}[t]
\includegraphics[height=15em]{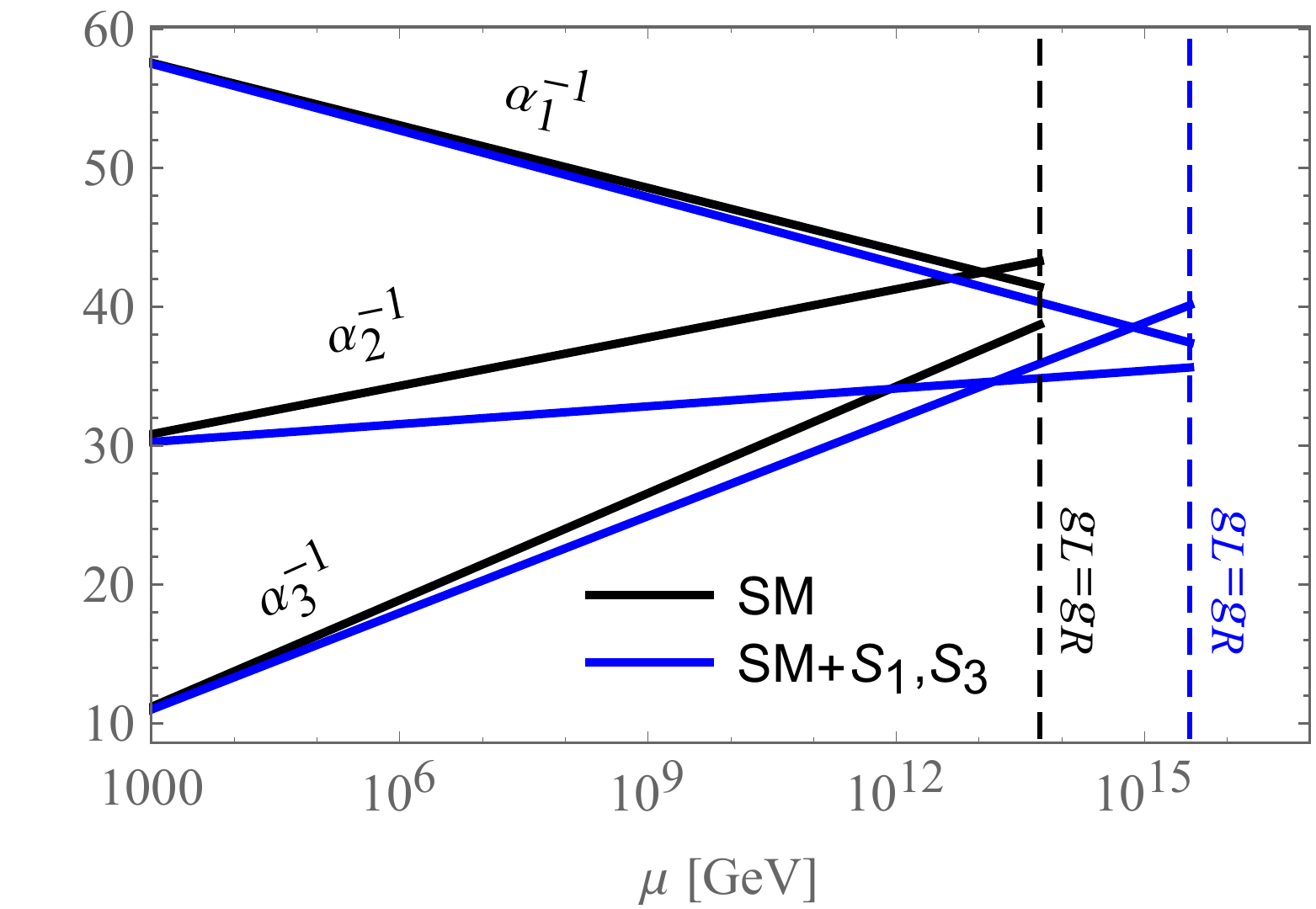}
\caption{Left--right gauge coupling unification in the SM and assuming only light additional $S_1$ and $S_3$ states. The dashed lines denote the scale at which the unification condition~\eqref{eq:unification}, $g_L=g_R$, is satisfied.}
\label{fig:running_gauge}
\end{figure}

Since two lines almost always meet at one point, Eq.~\eqref{eq:unification} can be generically satisfied even in presence of additional light particles beyond the SM, unless: (i) many new thresholds intervene that destroy the asymptotic freedom of the colour gauge coupling and make it grow faster than the hypercharge coupling; or (ii) the couplings would unify at non-perturbative values or beyond the Planck scale. Given that the running assuming only light SM degrees of freedom is rather far from these exceptional situations, we expect left--right unification to be generically possible. Since we do not presume to have knowledge of the full spectrum up to the PS scale, we simply show a proof of concept in Fig.~\ref{fig:running_gauge}, in which only the LQs required for the explanation of the anomalies lie below the PS scale. This shows that successful $g_L = g_R$ unification is achieved, at a scale slightly higher than in the SM. However, one should keep in mind that this should only be considered as an exercise because the presence of additional states in the \emph{desert}, especially if coming in generation copies, could change the unification scale by a few orders of magnitude.

Note that it is in general possible to decouple the spontaneous parity and PS breaking scales~\cite{Chang:1983fu} and thus have $g_R \neq g_L$ at $\mu_\text{PS}$ but still $y_L = y_R$ for the Yukawas, at least at tree level. This makes it possible to significantly lower the PS breaking scale, potentially down to the experimentally excluded value $\mu_\text{PS}\sim \unit[1000]{TeV}$~\cite{Volkas:1995yn}. While we do not pursue this possibility here, it should be kept in mind as an interesting alternative.

\bibliographystyle{utcaps_mod}
\bibliography{BIB}

\end{document}